\documentclass[12pt,preprint]{aastex}
\usepackage{emulateapj5,apjfonts}
\usepackage{natbib}
\usepackage{graphicx}
\newcommand{\be}{\begin{equation}}
\newcommand{\ba}{\begin{eqnarray}}
\newcommand{\ee}{\end{equation}}
\newcommand{\ea}{\end{eqnarray}}

\def\lesssim{\mathrel{\hbox{\rlap{\hbox{\lower4pt\hbox{$\sim$}}}\hbox{$<$}}}}
\def\gtrsim{\mathrel{\hbox{\rlap{\hbox{\lower4pt\hbox{$\sim$}}}\hbox{$>$}}}}

\def\gtsima{$\; \buildrel > \over \sim \;$}
\def\ltsima{$\; \buildrel < \over \sim \;$}
\def\gsim{\lower.5ex\hbox{\gtsima}}
\def\lsim{\lower.5ex\hbox{\ltsima}}
\def\simgt{\lower.5ex\hbox{\gtsima}}
\def\simlt{\lower.5ex\hbox{\ltsima}}
\def\simpr{\lower.5ex\hbox{\prosima}}

\def\ga{\gsim}

\def\simless{\mathbin{\lower 3pt\hbox
   {$\rlap{\raise 5pt\hbox{$\char'074$}}\mathchar''7218$}}}   
\def\simgreat{\mathbin{\lower 3pt\hbox
   {$\rlap{\raise 5pt\hbox{$\char'076$}}\mathchar''7218$}}}   

\begin{document}

\title{Relativistic Ionization Fronts}
\author{Paul~R.~Shapiro$^1$, Ilian~T.~Iliev$^2$, Marcelo A.~Alvarez$^1$,
 and Evan Scannapieco$^3$}

\altaffiltext{1}{Department of Astronomy, University of Texas, Austin, TX
  78712-1083, USA, shapiro@astro.as.utexas.edu}
\altaffiltext{2}{Canadian Institute for Theoretical Astrophysics, University
  of Toronto, 60 St. George Street, Toronto, ON M5S 3H8, Canada}
\altaffiltext{3}{Kavli Institute for Theoretical Physics, Kohn Hall, UC Santa
  Barbara, Santa Barbara, CA 93106, USA}

\label{firstpage}

\begin{abstract}

We derive the equations for the
propagation of relativistic ionization fronts (I-fronts)
in both static and moving gases, including
the cosmologically-expanding intergalactic medium (IGM).  We focus on
the supersonic R-type phase that occurs right after a source turns on, and
we compare the nonrelativistic and relativistic solutions for several
important cases.  Relativistic corrections can be significant up until
the light-crossing time of the  equilibrium Str\"omgren
sphere.  When $q,$ the ratio of this light-crossing time and the
recombination time, exceeds unity, the time for the expanding I-front
to reach the Str\"omgren radius is delayed by a factor of
$q$.    For a static medium, we obtain exact analytical solutions and
apply them to the illustrative problems of an O star in a molecular
cloud and a starburst in a high-redshift cosmological halo.
Relativistic corrections  can be important at early times when the H
II regions are small,  as well as at later times, if a density
gradient  causes the I-front to accelerate.  For the
cosmologically-expanding IGM, we derive an analytical solution in 
the case of a steady source and a constant clumping factor.   Here
relativistic corrections are significant for short-lived,
highly-luminous sources like QSOs  at the end of reionization
($z\simeq 6$), but negligible for weaker or higher-redshift
sources.   Finally, we numerically calculate the evolution of
relativistic I-fronts in the presence of small-scale structure and
infall, for a large galaxy undergoing a starburst and a luminous,
high-redshift QSO.  For such strong and short-lived ($z \sim 7$) sources,
the relativistic corrections are quite significant, and small-scale
structure can decrease the size of the H~II region by  up to an
additional $\sim 25$\%. However, most of the IGM was ionized  by smaller,
higher redshift sources.  Thus the impact of relativistic corrections
on global reionization is small and can usually be neglected.

\end{abstract}


\section{Introduction}

Ionization fronts are ubiquitous in astrophysics.  On interstellar
scales, they occur when new stars form in molecular clouds,
and their propagation has important dynamical effects on 
the surrounding medium. The compact and ultra-compact H~II regions that form
in this way have been extensively studied both theoretically and observationally 
\citep[for recent detailed reviews and further references see e.g.][]
{1999PASP..111.1049G,1999ARA&A..37..311E,2002ARA&A..40...27C}. 
Radiative feedback effects play a key
role in determining the physical conditions in star-forming regions, 
with profound effects on the star-formation rates and efficiencies, and the
resulting initial mass function. For example, ionization fronts may trigger 
star formation in dense clumps through compression and subsequent cooling of 
the gas 
\citep[e.g.][]{1997A&A...324..249L,2005ApJ...623..917H,2005ApJ...624..808L}.

On larger scales, the propagation of ionization fronts has emerged as
one of the key problems in cosmology.   \citet{1987ApJ...321L.107S}
pointed out that the first sources of ionizing  radiation in the
post-recombination expanding universe, be they quasars, primeval
galaxies, or some pregalactic objects like Population III stars, would
have caused cosmological I-fronts to sweep outward into the
surrounding intergalactic medium (IGM), and they derived the equations
to describe their propagation.   These I-fronts, they found, were
generally weak, R-type fronts that moved supersonically with respect
to both the neutral gas ahead of them  and the ionized gas behind,
thereby racing ahead of the hydrodynamical response of the ionized
gas. The H~II regions created by these expanding cosmological
I-fronts generally do not fill their ``Str\"omgren spheres'' -- i.e. a
fully-ionized sphere whose size is just large enough to encompass as
many atoms recombining per second as there are ionizing photons
emitted per second by the central source
\footnote{In view of this it would not be correct to refer, in
general, to cosmological H II regions as ``Str\"omgren spheres,''
since that term refers only to their equilibrium asymptotic size,
which they generally do not obtain.}.

The origin of this difference between the cosmological
I-fronts and their interstellar counterparts is the ever-decreasing
mean gas density in the cosmological case.  The predominance of this
R-type phase in the cosmological case has made it  possible to
approximate the propagation of the global I-fronts believed
responsible for the reionization of the universe by a ``static limit''
in  which the evolving inhomogeneous density field of the IGM is
assumed to be  predetermined by large-scale structure formation,
neglecting the  hydrodynamical response of the gas to its ionization
\citep[e.g.][]{2000MNRAS.314..611C,2003MNRAS.344..607S}.  On small
scales,  where gas can be much denser than  average, such as inside
halos, this ``static'' approximation can break down as the I-fronts
decelerate to about twice the sound speed of the ionized gas and
transform from supersonic R-type to subsonic D-type, thereby coupling
their fate to the dynamical response of the ionized gas.  Some attempt
has also been made to study this dynamical phase of cosmic
reionization I-fronts,  by ``zooming in'' on the gas dynamics of the
I-front as it encounters  individual halos
\citep*[e.g.][]{2004MNRAS.348..753S,2005MNRAS...361..405I} or by
relaxing the ``static approximation'' referred to above while
simulating large-scale structure formation during global reionization
\citep{2000ApJ...535..530G,2002ApJ...575...49R}. Here the primary
issue has been to characterize the onset, duration, and topology of
reionization
\citep*{2000ApJ...535..530G,2002MNRAS.330L..43B,2003ApJ...595....1H,
2003ApJ...586..693W,2003MNRAS.343.1101C,
2004ApJ...613....1F,2005ApJ...624..491I}, particularly in light of the
discrepancy between the epoch of reionization $z \gtrsim 15$ favored
by the first-year data from the {\em Wilkinson Microwave Anisotropy Probe}
\citep{2003ApJS..148..161K}, and observations of a Gunn-Peterson
trough in  high-redshift quasars, which suggest 
that reionization did
not end until  $z \sim 6$ \citep{2003AJ.126..1W,2004AJ....128..515F}.  
Other important questions include the growth of individual ionized regions
around high-redshift Lyman-alpha emitters
\citep{2002ApJ...576L...1H,2004MNRAS.349.1137S},   which would
otherwise be highly obscured by neutral gas
\citep{2004ApJ...617L...5M,2005ApJ...619...12S},  and the relationship
between the proximity effect around high-redshift quasars and the mean
IGM neutral fraction at this redshift
\citep{2004Natur.432..194W,2005ApJ...620L...9O,2005ApJ...620...31Y}.

Until recently, the I-fronts described above have been treated 
nonrelativistically, as if the speed of light were infinite.
Unlike shock fronts, which can reach relativistic (Rel) speeds only in the
most extreme environments where the energy density is very high, such that
the kinetic energy per baryon is comparable to or greater than its rest mass
energy, ionization fronts are
intrinsically radiative phenomena that are mediated by the propagation
of photons.  This means that even the weakest ionizing
source may, in principle, drive a strongly relativistic I-front.  
Thus the finite speed of light is {\em always} an
issue  for I-front propagation, whose importance should be ascertained
for each set  of astrophysical circumstances.

Nevertheless, relativistic I-fronts have received only limited
treatment in the literature.  Roughly speaking, the I-front will be
relativistic whenever the photon to atom ratio at the front exceeds
unity.  This occurs whenever a source releases ionizing photons faster
than a surface moving outward at the speed of light can overtake as
many neutral atoms.  The equations of \citet{1987ApJ...321L.107S} for
the evolution of the spherical I-front around a point source in the
cosmologically-expanding IGM show that, at early times close to source
turn-on, the nonrelativistic (NR) treatment formally yields an I-front
speed which exceeds the speed of light, for just this reason.   This
``superluminal'' phase is generally so short-lived, lasting only a
tiny fraction of the age of the  universe at source turn-on, that it
has a negligible effect on the overall evolution of the I-front for a
steady source. Recently, however, interest in this brief, early  phase
of the I-front has been drawn by observations of the spectra of high
redshift quasars and Lyman alpha line emitters at redshifts $z>6$
which  suggest that we are seeing each of these sources as if it is
surrounded by an isolated intergalactic H II region of its own
making.  In the quasar case, in particular, assumptions that  the
source of ionizing radiation is the observed quasar, itself, and that
quasar lifetimes are much less than the age of the universe at those
epochs,  combine to suggest that the relativistic phase of the I-front
predicted by  the nonrelativistic solutions is quite relevant.  For
this reason, \citet{2003AJ.126..1W} presented a  revised equation for
the expansion speed of the spherical I-front which takes  account of
the finite speed of light, although it neglects recombinations.   They
noted, however, that the relationship between the apparent size of the
H~II region along the line of sight and the parameters on which the
physical  evolution of that size depends, namely the quasar's
luminosity and the average density of neutral H atoms in the IGM
outside the H~II region, are exactly the same as  would be achieved by
applying the equations of \citet{1987ApJ...321L.107S}.  This
surprising result is apparent also in the treatments of
\citet{2005ApJ...623..683Y} and \citet{2005ApJ...620...31Y}, who
discussed the  effect of relativistic expansion on the apparent shape
of a cosmological H II region around a high-redshift QSO, including
the effects of recombination.  It results from the fact that the
photons in the observed quasar spectrum at the redshifted wavelength
which marks the outer edge of the H II region were emitted by the QSO
at the {\em retarded} time equal to the time those photons reached the
I-front, minus their  light-travel time from the source.  Using this
{\rm retarded} time to evaluate the I-front radius in the equations of
\citet{1987ApJ...321L.107S} automatically accounts properly for the
finite speed of light.

These recent discussions of the QSO H II regions state their basic H
II radius evolution equation without derivation and do not explain its
connection to the fundamental conservation equations for an I-front,
namely, the I-front continuity jump condition and the equation of
radiative transfer for the  ionizing photons.  In what follows, we
will start from those fundamental  conservation equations to take full
account of special relativity in  generalizing them to address a wider
range of circumstances and make clear the derivation that underlies
the evolution equation for the spherically symmetric H~II regions
mentioned above.  We will examine the general problem  of relativistic
I-fronts, in both static, moving, and cosmologically-expanding media,
deriving a number of exact and universal solutions and assessing the
importance of relativistic effects in both the interstellar medium and
cosmology.

The structure of this work is as follows. The relativistic equations
for I-front  evolution in a general form are derived in
\S~\ref{general_sect} for a source in a moving gas. In
\S~\ref{noncosmo_sect} we use these equations to describe the
relativistic evolution of a spherical  I-front surrounding a point
source in a static medium, deriving some exact analytic solutions to
this problem, for a uniform  gas density and a density that  varies as
a power-law in radius.  For illustration, we apply these to a front
expanding around an O-star or OB association in a molecular cloud and
to a front inside a cosmological halo.  We also present a  numerical
solution for the I-front around a point source in a plane-stratified
medium. In \S~\ref{cosmo_sect} we generalize the equations of
\citet{1987ApJ...321L.107S} for the cosmological I-front around a
point source in an expanding IGM to take account of special
relativity, finding an exact analytical solution, as well, which we
use to assess the importance of relativistic effects as a function  of
redshift and  source flux. In \S~\ref{MH_sect}, we  present numerical
solutions for the isolated H~II regions around cosmological sources,
which include the effect of small-scale cosmological structure
(i.e. self-shielding minihalos, time-varying clumping factor for the
diffuse IGM, and the infall of gas surrounding each source halo) on
these intergalactic I-fronts. We apply these solutions to  determine
the impact of relativistic effects on global cosmic reionization, by
the approach described in \citet{2005ApJ...624..491I}, which evolves
the I-fronts created by a statistical distribution of cosmological
ionization sources to the point at which their H II regions finally
overlap. Our conclusions are summarized in \S~\ref{summary_sect}.

\section{Basic Equations}
\label{general_sect}
\subsection{Relativistic I-front in a Moving Gas}
Consider the motion of an I-front through some gas that is itself moving with
respect to the source of the ionizing radiation.  In the lab frame defined as
the rest frame of the source, the I-front moves with velocity 
${\bf v}_I\equiv {\bf \beta}_Ic$, while the gas just outside the I-front on
the neutral side (side ``1'') moves with velocity ${\bf v}_{g,1}\equiv 
{\bf \beta}_{g,1}c$.
Assume that ${\bf v}_I$ and ${\bf v}_{g,1}$ are in the
same direction, so the motion is one-dimensional.  The relativistic law for the
addition of velocities then gives the velocity of the I-front as measured in 
the rest frame of the gas just outside the front, $v_I'=\beta_I'c$,
according to 
\be
v_I'=\frac{v_I-v_{g,1}}{1-v_Iv_{g,1}/c^2}.
\label{app1}
\ee
In the rest frame of the I-front, the gas moves with velocity $v_{g,1}''=
\beta''_{g,1}c=-v_I'=-\beta_I'c,$
where here and below double primed variables are quantities measured 
in the frame of the front.  

\subsection{I-front Continuity Jump Condition}
The relativistic I-front continuity jump condition, which expresses the balance
between the flux of ionizing photons arriving at the front and the flux of 
atoms that are ionized by those photons as the atoms pass through the front, is
written in the rest frame of the I-front as follows:
\be
n_{H,1}''v_{g,1}''=\beta_i^{-1}F'',
\label{app6}
\ee
where $v_{g,1}''$ is the front velocity relative to the 
undisturbed gas into which the front is moving, given by $v_{I,1}'$ in equation
(\ref{app1}).
To get the density $n_{H,1}''$, we take the value measured in the rest frame
of the I-front, 
\be
n_{H,1}''=\gamma_{g,1}''n_{H,1}'=\gamma_I'n_{H,1}',
\label{app7}
\ee
where the Lorentz factors, $\gamma$, are related to the velocities, $\beta c$,
by $\gamma\equiv (1-\beta^2)^{-1}$, as usual.  The quantity $\beta_i$ counts 
the number of ionizing photons required to yield a single newly ionized H atom 
on the ionized side of the front and is thus frame-independent.

The ionizing photon flux $F''$ as measured by the moving front is different 
from the lab-frame flux $F$ measured in the frame of the source, by a 
Lorentz-boost to the frame moving with velocity $v_I$ in the lab frame:
\be
F''=\gamma_I(1-\beta_I)F
\label{app8}
\ee
\citep*{1986ApJS...60..393S}, where we have implicitly assumed that the
relativistic Doppler shift of the photons from the lab frame to the rest frame
of the moving gas does not significantly affect the integrated number of 
ionizing photons in the incident flux.  Otherwise, we should replace $F$ in 
equation (\ref{app8}) by the modified flux which counts only those photons at
the location of the I-front at lab frame frequencies $\nu$ such that $\nu \gg
\nu_{min}$, where 
\be
\nu_{min}=\nu_{th}\gamma_{g,1}(1+\beta_{g,1}),
\label{app9}
\ee
where $\nu_{th}$ is the threshold frequency for ionizing $H$ atoms from the 
ground state.  

Combining equations (\ref{app1}) and (\ref{app6})-(\ref{app8}) yields the 
equation for the I-front velocity in the lab frame as follows:
\be
\beta_I=\left[\frac{(1-\beta_I\beta_{g,1})\gamma_{g,1}}
{\gamma_I'}\right]\left[\frac{\gamma_I(1-\beta_I)F}{\beta_in_{H,1}c}\right]+
\beta_{g,1}
\label{app10}
\ee
We have also used the fact that the lab frame $H$ atom density $n_{H,1}$ is
related to the value in the rest frame of the gas, according to 
\be
n_{H,1}=n_{H,1}'\gamma_{g,1}.
\label{app11}
\ee

\subsection{Radiative Transfer of Ionizing Flux Across H II Region}
\subsubsection{Optical Depth}
Ionizing photons traveling radially outward from the central source will 
suffer absorptions due to the residual bound-free opacity of gas on the ionized
side of the I-front, inside the H II region.  As these photons cross a 
spherical shell of radius $dr$ as measured in the lab frame (i.e. rest frame
of source), in which frame the H atom number density is $n_H(r)$ and the 
frame-independent neutral fraction is $(1-x)$, the probability that a photon
emitted at a frequency $\nu$ in the lab frame is absorbed is given for a shell 
that expands away from the source at velocity $v_g(r)=\beta_g(r)c$, by optical
depth $d\tau$, according to:
\be
d\tau_{\nu}=n_H(1-x)\sigma_{\nu'}(1-\beta_g)dr,
\label{app12}
\ee
where $\sigma_{\nu'}$ is the photoionization cross section of an H atom in the
ground state, evaluated at photon frequency $\nu'=\gamma_g(1-\beta_g)\nu$, the
frequency in the comoving frame of the expanding shell
\citep{1986ApJS...60..393S}.

\subsubsection{Ionization Equilibrium}
For gas in the expanding shell, ionization equilibrium
in the ``on-the-spot'' approximation, in which radiative recombinations to
atomic levels $n\ge 2$ are balanced by photoionizations from the ground state,
gives the neutral fraction according to
\be
\frac{1-x}{x^2}=\frac{\alpha_B(n_H')^2}{n_H'\overline{\sigma}F'}=
\frac{\alpha_Bn_H'}{\overline{\sigma}F'},
\label{app13}
\ee
where $\overline{\sigma}$ is the frequency-averaged bound-free cross-section
for frequencies $\nu'\ge \nu_{th}$, weighted by the flux of photons at each 
frequency, and where the primed quantities are measured in the comoving frame
of the gas. The atomic number density and the photon flux in this frame are
related to their counterparts in the lab frame according to 
\be
n_H'=n_H/\gamma_g,
\label{app14}
\ee
and
\be
F'=\gamma_g(1-\beta_g)F,
\label{app15}
\ee
where $F$ should be the photon flux for $\nu\ge\gamma_g(1+\beta_g)\nu_{th}$
\citep{1986ApJS...60..393S}.  Equations (\ref{app13})-(\ref{app15}) yield
\be
\frac{1-x}{x^2}=\frac{\alpha_Bn_H}{\overline{\sigma}\gamma_g^2(1-\beta_g)F}.
\label{app16}
\ee
According to equations (\ref{app12}) and (\ref{app16}), the optical depth
through the shell is given, therefore, by 
\be
d\tau_\nu=\left(\frac{\sigma_{\nu'}}{\overline{\sigma}}\right)
\left(\frac{\alpha_Bx^2n_H^2}{\gamma_g^2F}\right)dr.
\label{app17}
\ee

\subsubsection{Flux Variation With Distance From Source}
Consider the lab frame quantity $S(r,t_R)=4\pi r^2F(r,t)$, which is
the rate at which ionizing photons pass through a shell of radius
$r$ at time $t$ that were emitted by the source at the retarded time
$t_R\equiv t-r/c$. 
The quantity $S(0,t)$ is the ionizing photon luminosity of the source at time $t$. If the
source luminosity is $L_\nu\, (\rm erg\,s^{-1}Hz^{-1})$ in the lab frame, then
\be
S(0)=\dot{N}_\gamma\equiv\int_{\nu_{\rm th}}^{\infty}\dot{N}_{\gamma,\nu}d\nu
\equiv\int_{\nu_{\rm th}}^{\infty}\frac{L_\nu}{h\nu}d\nu.
\label{phot_lum}
\ee
In the absence of opacity due to photoionization, the quantity $S(r,t_R)$ is 
independent of $r$. The effect of opacity is to attenuate $S(r,t_R)$ as
follows, using equation (\ref{app17}):
\ba
\frac{d}{dr}\left[4\pi r^2F_\nu(r)\right]&=&
-\dot{N}_{\gamma,\nu}(t_R)e^{-\tau_\nu}\frac{d\tau_\nu}{dr}\nonumber\\
&=&-\left[4\pi r^2F_\nu(r)\right]\left[\left(\frac{\sigma_{\nu'}}
{\overline{\sigma}}\right)\left(\frac{\alpha_Bx^2n_H^2}{\gamma_g^2F}
\right)\right].
\label{app18}
\ea
Multiplying both sides of equation (\ref{app18}) by $d\nu$ and integrating over
all frequencies $\nu\ge\gamma_g(1+\beta_g)\nu_{th}$ yields
\be
\frac{dS}{dr}=-4\pi r^2\alpha_Bx^2n_H^2\gamma_g^{-2}.
\label{app19}
\ee
We can integrate this over radius to yield $S(r_I,t_{R,I})$ at the location of
the I-front according to  
\be
S(r_I,t_{R,I})=S(0,t_{R,I})-\int_0^{r_I}dr\cdot4\pi r^2xn_en_H\gamma_g^{-2}\alpha_B,
\label{app20}
\ee
where $t_{R,I}\equiv t_R(r=r_I)$ and we evaluate the quantities inside the
integral on the right-hand-side at the time $t(r)\equiv t_{R,I}+r/c$. 
We have also replaced one of the factors of $xn_H$ by $n_e$ above. Although we
have neglected the presence of helium until now, we can account for the
contributions of ionized helium to the electron density inside the H~II region
by writing $n_e=\chi_{\rm eff}xn_H$, where $\chi_{\rm eff}\equiv 1+p A({\rm
  He})$, $A({\rm He})=0.08$ is the He abundance by number with respect to
hydrogen, and $p=0,1$ or 2 according to whether He is mostly neutral, singly
ionized or doubly ionized, respectively. If we also allow for a
volume-averaged clumping factor, $C=\langle n_H^2\rangle/\langle n_H\rangle^2$,
then equation (\ref{app20}) 
becomes 
\be
S(r_I,t_{R,I})=\dot{N}_\gamma(t_{R,I})-\int_0^{r_I}dr\cdot 4\pi
r^2\chi_{\rm eff}x^2n_H^2\gamma_g^{-2} \alpha_BC.
\label{app21}
\ee
Here $\alpha_B = 2.59 \times 10^{-13}$ cm$^3$ s$^{-1}$ is the case B 
recombination coefficient for hydrogen at $10^4$ K (which is the temperature
we assume for the ionized medium throughout this paper). 
The value of the radius $r_I$ which makes $S(r_I,t_{R,I})$ vanish according to
equation~(\ref{app21}) is then the Str\"omgren radius $r_S$. Hereafter, 
unless otherwise
noted, the only retarded time to which we shall refer will be $t_{R,I}$, and
we shall drop the subscript ``I'' when referring to it. The lab frame flux at
the front at time $t$ is then given by 
\be
F(r_I,t)=\frac{S(r_I,t_R)}{4\pi r_I^2}.
\label{flux_at_front}
\ee 

While the analysis in \S~\ref{general_sect} apparently applies to a
point source of ionizing radiation surrounded by a
spherically-symmetric gas, it actually applies equally well to the
general case of inhomogeneous gas which varies in all three spatial
dimensions. In the latter case, we simply interpret the equations as
if they are written in $(r,\theta,\phi)$ spherical coordinates
centered on the source, as long as any non-radial gas motions which
might be present are subrelativistic. If a problem arises in which
relativistic,  non-radial gas motions occur, the equations above
should be appropriately modified to account for the effect of
transverse velocity on the relativistic transformations, as well. In
the absence of such relativistic transverse gas motion, however, we
can use the analysis presented in this section for the general 3D
problem without modification.

\section{Relativistic I-fronts in Static Media} 
\label{noncosmo_sect}

\subsection{Point Source in a Static Medium}
\label{static_subsec}

Consider a point source in a static gas. In this case, $v_{g,1}=0$,
the velocity of the I-front $v_I'$ in the frame of the undisturbed gas
ahead of the front is just the same as its velocity $v_I$ in the lab
frame in which the source is at rest, and $\gamma_I'=\gamma_I$. In the
frame of the front, therefore, $v_{g,1}''=-v_I$, the H atom density
$n_{H,1}''=\gamma_In_{H,1}$, and the flux $F''$ is related to the lab
frame flux $F$ by equation~(\ref{app8}).  Equation~(\ref{app10}) for
the front velocity in the lab frame then becomes  \be
\beta_I=\frac{F}{\beta_in_{H,1}c+F},
\label{I-front_vel}
\ee  where $F$ is the number flux of ionizing photons at the current
position of the I-front, given by equation~(\ref{app21}) with
$\gamma_g=1$ and equation~(\ref{flux_at_front}).

Combining equations~(\ref{app21}) - (\ref{I-front_vel}) and solving
for  ${\rm v}_{I}$, we obtain the lab frame evolution equation for the
radius $r_I$ of the spherical I-front, \be \frac{dr_I}{dt}
=c\frac{\dot N_\gamma(t_R)- 4\pi\int_0^{r_I}{n_H}^2 C\alpha_B\chi_{\rm
eff}r^2dr} {4\pi \beta_icr_I^2n_{H,1}+\dot N_\gamma(t_R)- 4\pi
\int_0^{r_I}{n_H}^2 C\alpha_B\chi_{\rm eff}r^2dr}.
\label{I-front}
\ee  In the nonrelativistic limit in which $dr_{I}/dt \ll c$, the
first term in the denominator becomes dominant and
equation~(\ref{I-front}) reduces to the  usual expression  \be
\frac{dr_I}{dt} =\frac{\dot N_\gamma(t)- 4\pi \int_0^{r_I}{n_H}^2
C\alpha_B\chi_{\rm eff}r^2dr} {4\pi \beta_ir_I^2n_{H,1}}.
\label{I-front_nonrel}
\ee  In the opposite, strongly relativistic limit, in which  the flux
is very large, both the denominator and numerator are dominated by the
$\dot N_\gamma$ term and $dr_I/dt$ approaches  $c$ asymptotically.

For this static medium, the I-front velocity equation can also be
derived by imposing photon conservation within the ionized volume.
This is done by requiring that at a time $t$ the total number of
photons that have reached the ionization front is equal to the sum of
the total number of atoms within the volume and  the total number of
recombinations that occurred within the volume {\em excluding}
recombinations that affect emitted photons that have not yet reached
$r_I(t)$, that is \ba
\int_0^{t_R}\dot{N}_\gamma(t')dt'&=&4\pi\beta_i\int_0^{r_I(t)}n_H(r)r^2dr\nonumber\\
&+&4\pi\int_0^{t}dt'\int_0^{r_I(t')}n_H^2(r)C\alpha_B\chi_{\rm
eff}r^2dr\nonumber \\
&-&4\pi\int_{0}^{t-t_R}dt'\int_0^{ct'}n_H^2(r)C\alpha_B\chi_{\rm
eff}r^2dr
\label{eq:global_conserv}
\ea 
Differentiating this equation with respect to time, we obtain 
\ba
\left[\dot{N}_\gamma(t_R)-4\pi \int_0^{r_I}n_H^2(r) C\alpha_B\chi_{\rm
eff}r^2dr\right] \left(1-\frac{\dot{r_I}}{c}\right)\nonumber \\=
4\pi\beta_i\frac{dr_I}{dt}n_H(r_I)r_I^2,   
\ea and solving for
$\dot{r}_I$ recovers equation~(\ref{I-front}).

For a given density field and time-dependent photon luminosity,
equation~(\ref{I-front}) can, in principle, be solved to yield the
I-front radius and radial velocity as explicit functions of time $t$,
for any direction $(\theta,\phi)$. We will do so for certain density
profiles in the sections that follow. Before we consider such explicit
solutions, however, we note that there is a simplification that
results if we rewrite our equations in terms of the retarded time,
$t_R$, rather than physical time, $t.$  For every interval $dt$ of
physical time, there is a corresponding interval  \be
dt_R=dt(1-\beta_I).
\label{dt_R}
\ee If we define a ``retarded velocity'' $\beta_{\rm I,R}\equiv
c^{-1}dr_I/dt_R$, then equation~(\ref{dt_R}) says that $\beta_{\rm
I,R}$ is related to $\beta_I$ according to  \be \beta_{\rm
I,R}=\frac{\beta_I(t)}{1-\beta_I(t)}.
\label{beta_RI}
\ee   Combining equations~(\ref{flux_at_front}), (\ref{I-front_vel})
and (\ref{beta_RI}) then yields an equation for $dr_I/dt_R$ in terms
only of $t_R$, as follows: \be
\frac{dr_I}{dt_R}=\frac{S(r_I,t_R)}{4\pi r_I^2n_{\rm H,1}\beta_i}.
\label{v_R}
\ee Remarkably enough, equation~(\ref{v_R}) is identical to the
nonrelativistic equation for the I-front velocity $dr_I/dt$,
equation~(\ref{I-front_nonrel}), if we replace $t$ in the latter
equation by $t_R$. \footnote{A similar correspondence was noted by Yu (2005)
(see equation [7] there).}
Therefore, any solution of equation~(\ref{v_R}) for
the explicit dependence of $r_I$, and, using equation~(\ref{beta_RI}),
of $\beta_I$, on $t_R$ becomes an implicit solution for the dependence
of $r_I$ and $\beta_I$ on physical time $t$ by relating $t$ to $t_R$
according to  \be t=t_R+r_I(t_R)/c.
\label{t_tR}
\ee This universal property will allow us to simplify the solution of
the I-front evolution equation and yield analytical results when the
explicit equation~(\ref{I-front}) otherwise would seem to require a
numerical solution.

If we specialize to the case in which the source luminosity after
turn-on, $\dot{N}_\gamma$, is time-independent, then, according to
equations~(\ref{app21}) and (\ref{v_R}), the ``retarded velocity''
$\beta_{I,R}$, when the I-front radius has some value $r_I$ is
identical to the velocity $v_{\rm I,NR}$ of the nonrelativistic
I-front solution at the same radius, divided by $c$. In that case, we
can write the dependence of the relativistic I-front velocity
$v_I=c\beta_I$ on $r_I$, using equations~(\ref{beta_RI}) and
(\ref{v_R}), directly in terms only of the nonrelativistic solution at
the same radius, according to  \be
v_I(r_I)=\frac{v_{I,NR}(r_I)}{1+v_{I,NR}(r_I)/c},
\label{v_I_from_NR}
\ee regardless of the difference in physical times at which the
I-fronts in the relativistic and the nonrelativistic solutions achieve
the radius $r_I$.  

The relativistic I-front solution for the radius versus time,
which depends on the source luminosity and the atomic density in the
external medium, describes the size of the H~II region in the
source frame.  In order to use this solution to interpret an
observed H~II region, however, we must take account of the fact that
photons which travel from the I-front to the observer have different
travel times if they originate from different locations along the
front surface but arrive at the same time.  Photons received from the
part of
the I-front which moves towards us along the line of sight to the
central source travel a smaller distance to reach us
than did photons arriving at the same time from the part of the I-front
which moves transverse to that line of sight.  As a result, the
line-of-sight radius will appear to be bigger than the transverse
radius, since we see it as it was, later in the time evolution of the
spherical front.  Suppose we use the apparent transverse size measured
at the distance of the source to fix the time $t$ (measured from source
turn-on) at which the relativistic I-front solution matches
this radius, $r_I(t)$.  In that case, the apparent line-of-sight radius will
be given, instead, by $r_{I,NR}(t)$, the solution of the nonrelativistic
I-front equations, evaluated at the same time $t$, since the time $t$
is then equal to the retarded time for the apparent radius of the I-front
along the line of sight.   This apparent departure of the shape of
the I-front from spherical was also discussed in the context of
quasar H II regions at high redshift by Yu (2005).
A related argument also explains why the nonrelativistic
solution for the I-front radius is valid for interpreting the
apparent line-of-sight sizes of the H II regions around high redshift quasars
based upon the
absorption spectra of these quasars, as mentioned in our introduction.

\subsection{Steady Source in a Uniform Density Gas}
\label{static_subsec1}
  If we assume a steady ionizing source that turns on in a gas with
 uniform density, clumping factor, and $\chi_{\rm eff}$, all
 independent of time, then  equation~(\ref{I-front}) has a closed-form
 analytical solution, which is obtained as follows. Let us define the
 dimensionless ionized volume as \be y\equiv V_I/V_S = V_I
 C\alpha_Bn_H^2\chi_{\rm eff} \dot{N}_\gamma^{-1}, \ee  where $V_S$ is
 the the Str\"omgren volume, and the dimensionless time as \be w\equiv
 t/t_{rec}=t\,C\alpha_B n_H\chi_{\rm eff}, \ee where $t_{rec}$ is the
 recombination time.  The non-dimensional form of  (\ref{I-front}) is
 thus given by \be \frac{dy}{dw}= \frac{(1-y)}{1+q(1-y)/(3y^{2/3})},
\label{dimrel}
\ee where  \be q\equiv \frac{r_S}{ct_{rec}}, \ee and $(4 \pi/3 )r_S^3
\equiv V_S$, i.e. $q$ is the light-crossing time  of the Str\"omgren
radius in units of the recombination time.  This equation reduces to
the  nonrelativistic one when $c\rightarrow\infty$, $q\rightarrow 0$,
\be \frac{dy}{dw}={1-y}, \ee which has the standard solution \be
y=1-e^{-w}.
\label{strom_nonrel}
\ee In general,  the solution of equation (\ref{dimrel}) is given by
the implicit relation \be w=qy^{1/3}-\ln(1-y).
\label{trans}
\ee The reader can easily confirm that this relativistic I-front
solution in equation~(\ref{trans}) is identical to that which results
if we replace $t$ in equation~(\ref{strom_nonrel}), the solution of
the nonrelativistic I-front equations, by the retarded time $t_R$, as
described in \S~\ref{static_subsec}.

\begin{figure*}
\includegraphics[width=\textwidth]{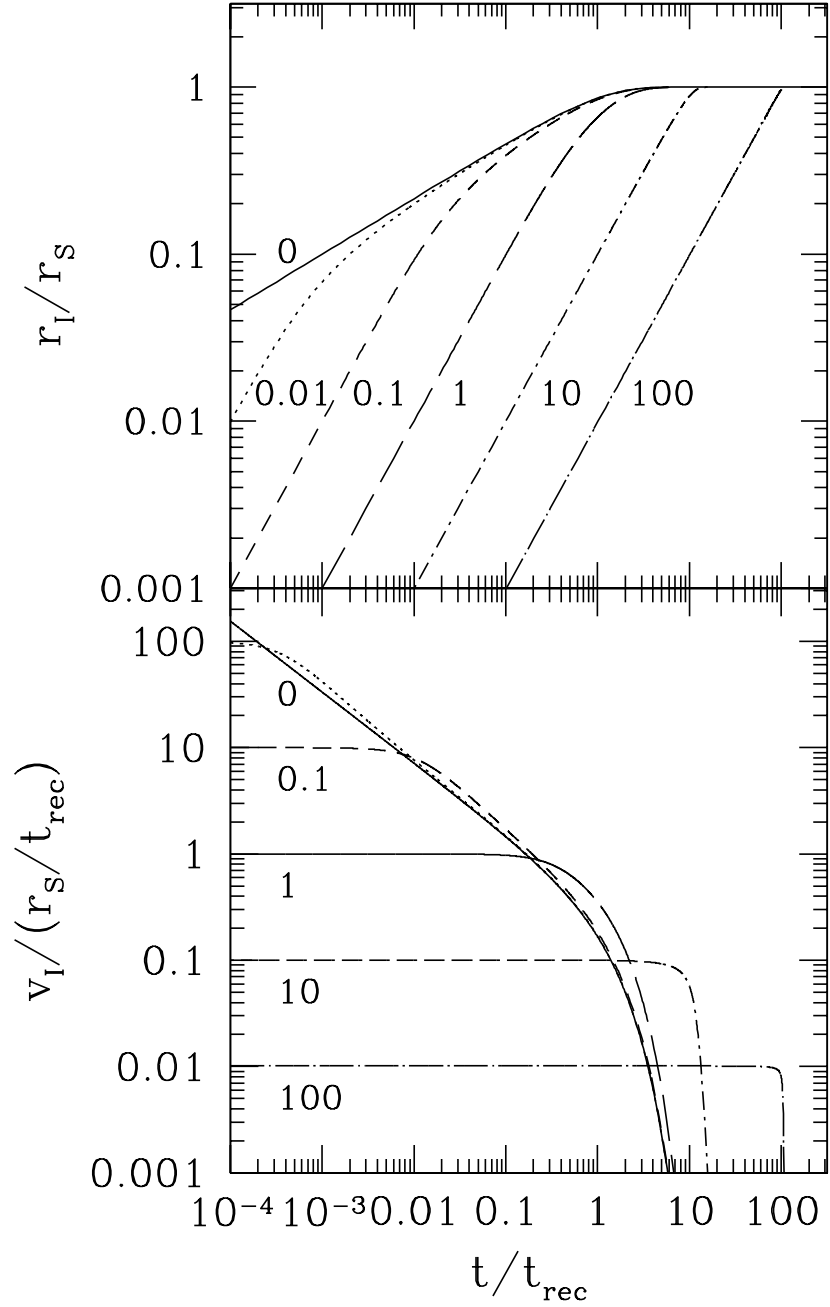}
\caption{ Relativistic I-front for a steady source in a static,
uniform gas: (a) (top) radius (in units of Str\"omgren radius $r_{\rm
S}$) and (b) (bottom) velocity (in units of $r_{\rm S}/t_{\rm rec}$).
Curves are labeled by values of the dimensionless light-crossing time
of the Str\"omgren radius, $q\equiv  r_{\rm S}/(ct_{\rm rec})$, with
$q=0$ (i.e. nonrelativistic limit) and  $q=0.01,0.1,1,10,$ and 100. In
these dimensionless units, the speed of light is $q^{-1}$.
\label{nondim_soln_fig}
}
\end{figure*}

In Figure~\ref{nondim_soln_fig} we show the I-front radius $r_I$ and
velocity $v_I$ versus time $t$ for different values of $q$.
When $q=0$, we recover the NR solution, in which the I-front starts
with  infinite velocity and decelerates to zero velocity within of
order the H~II region recombination time, by which time the I-front
radius has grown to equal the radius of the Str\"omgren sphere. When
$q>0$, the speed of  the I-front  remains close to the speed of light
from the time of source turn-on to roughly a time $qt_{\rm rec}$
later. For $q>1$, the I-front does not reach the size of the
Str\"omgren sphere until a time $qt_{\rm rec}$ after the source
turn-on.  In fiducial units, $q$ is given by  \be
q=0.019\,C^{2/3}{\dot{N}_{\gamma,49}}^{1/3}{n_{H,3}}^{1/3},
\label{q_fiducial}
\ee (assuming $\chi_{\rm eff}=1.08$, i.e. He is once-ionized)  where
$\dot{N}_{\gamma,49}=\dot{N}_{\gamma}/(10^{49}\,\rm s^{-1})$ and
$n_{H,3}=n_{H}/(10^3\,\rm cm^{-3})$.  This estimate shows that the
astrophysically-relevant values of $q$ can be significantly different
from 0. Hence, the relativistic correction to the standard I-front
propagation solution can be important,  especially in the immediate
vicinity of the source. The correction is even  more important when
the ionizing source is very luminous or the surrounding gas is dense
and highly clumped. In such strongly relativistic cases, the
correction would be  important during the entire R-type phase of the
evolution of the H~II region,  until the corresponding Str\"omgren
sphere is reached  (see Figure~\ref{nondim_soln_fig}).

To estimate the impact of our results for compact H~II regions  in
more detail, we consider an O-type star in a molecular cloud
environment.  Here the typical gas number densities are
$n_H\sim10^2-10^3\,\rm cm^{-3}$, but can reach $\sim10^4-10^5\,\rm
cm^{-3}$ in denser regions and up to $n_H\sim10^7\,\rm cm^{-3}$ in
so-called ``hot cores''
\citep{1999PASP..111.1049G,1999ARA&A..37..311E,2002ARA&A..40...27C}.
Furthermore,  the observed gas distribution is very inhomogeneous
\citep{1999PASP..111.1049G,1999ARA&A..37..311E,2001ApJ...560..806L},
and the filling factors of dense clumps are $f_{\rm v}=0.03-0.3$
\citep{1999ARA&A..37..311E}, roughly corresponding to clumping factors
$C\approx f_{\rm v}^{-1}=3-30$. Thus, in our estimates below we assume
the  mid-range value of $C=10$. Finally, for a typical massive O-type
star  the ionizing  photon flux is $\dot{N}_\gamma\sim10^{49}\,\rm
s^{-1}$, and up to a few times $10^{50}\,\rm s^{-1}$ for a cluster of
such stars with corresponding H~II region radii ranging roughly from
0.01 pc to $\sim 1$ pc \citep{1999PASP..111.1049G}.

How important are relativistic effects in such an environment? For
$C=10$,  $\dot{N}_\gamma\sim10^{49}\,\rm s^{-1}$, and
$n_H\sim10^5\,\rm cm^{-3}$, we obtain $q=0.41$, while using
$\dot{N}_\gamma\sim3\times10^{50}\,\rm s^{-1}$, and $n_H\sim10^4\,\rm
cm^{-3}$ results in $q=0.59$. As seen in Figure~\ref{nondim_soln_fig},
this means that the I-front expansion is relativistic all the way out
to the Str\"omgren radius, which means that the resulting H~II region
size during this rapidly-expanding R-type phase is much smaller than
the one predicted by the nonrelativistic expression.

On the other hand, this rapid expansion phase is completed within a
few recombination times or less, where $t_{\rm
rec}=1/(C\alpha_Bn_H\chi_{\rm eff})\lesssim1\,\rm yr$,  which is quite
short compared to the dynamical time of such compact H~II regions,
$t_{\rm dyn}\sim 5\times10^3$ yr \citep{1999PASP..111.1049G}.  Hence
most of their evolution is spent in the slow D-type phase, in which
the hot, over-pressured ionized bubble expands into the surrounding
medium, following the initial, supersonic  R-type phase.  During this
longer-lived D-type phase, the I-front moves with speed $v_I\ll c$,
and the  relativistic corrections are negligible. Numerically, in
terms of the H~II region final equilibrium  size, $\lesssim 5$\% of
the expansion  occurs during the initial relativistic phase.

Our approach  neglects the finite rise time of the source luminosity.
Yet, for the high densities typical of star-forming regions in
molecular clouds, this is much longer than the recombination time.
For a 30  $M_\sun$ star, for example, model star calculations find
that the UV luminosity beyond the Lyman limit doubles in 4000 years as
the star approaches the main sequence \citep{1969ARA&A...7...67M}.  In
this case, there may not be a  distinct relativistic R-type phase as
estimated above.  If the source luminosity $\dot{N}_\gamma$ increases
gradually over a time which is long compared with the recombination
time, then the dimensionless parameter $q$ defined above must also
increase gradually over time, in proportion to $\dot{N}_\gamma^{1/3}$,
as does the instantaneous Str\"omgren radius.  As such, the criterion
that $q$ is significantly larger than zero, necessary for relativistic
corrections to be important during the entire R-type phase, may not
occur before the I-front fills this gradually increasing Str\"omgren
radius.   On the other hand, realistic density profiles for the
immediate surroundings of a massive star when it turns on are likely
to be  far from uniform.  If there is a steep enough negative density
gradient  outside the star, then the I-front can accelerate outward
even if it initially decelerated and entered the D-type phase (Franco,
Tenorio-Tagle, and  Bodenheimer 1990).  This phenomenon is associated
with the ``champagne phase''  of the H~II region.  When this happens,
the I-front moves ahead of the  dynamical expansion of the inner H~II
region as a weak R-type front and can accelerate to relativistic
speeds again.  In that case, the finite rise time of the stellar
luminosity is not relevant, since the flux at the position of  the
I-front effectively rises in the time it takes light to cross a few
absorption mean free paths in the neutral gas outside the front.  We
will  discuss the propagation of relativistic I-fronts in density
gradients in \S3.3 and \S3.4 below.

\subsection{Steady-Source in a Power-law Density Profile}
\label{power-law_sect}

Consider a source that switches on in the center of a
spherically-symmetric  density profile $n(r)$. The case in which the
density decreases with increasing radius is relevant whenever the
source, like a massive star, is forming by gravitational instability,
in the middle of a centrally-concentrated gas profile.  As in
\S~\ref{static_subsec1}, we shall assume that the source luminosity is
time-independent and that the clumping factor is constant in space and
time. The nonrelativistic problem of an H~II region in a
spherically-symmetric density distribution that varies as a power of
radius outside of a flat-density core was discussed by
\citet{1990ApJ...349..126F}.  Let the H atom number density in the
undisturbed gas be defined by  \be n_H(r)=\left\{
\begin{array}{ll}
n_0(r/r_0)^{-\gamma}, & r>r_0,\\ n_0, & r\leq r_0.
\end{array}
\right.
\label{powerlaw}
\ee As long as the I-front is inside the core, its propagation follows
the solution for a uniform-density gas  derived in
\S~\ref{static_subsec1} above.  In this phase the I-front continually
slows down from $v_I\approx c$ at small radii $r_I$. If the core
Str\"omgren radius $r_{S,0}\leq r_0$, where
$r_{S,0}\equiv[3\dot{N}_{\gamma}/(4\pi C\alpha_Bn_0^2)]^{1/3}\leq
r_0$, then the I-front will slow down to zero velocity just as it
fills this Str\"omgren sphere, thereby remaining trapped within the
core. If $r_{S,0}>r_0$, instead, then the I-front continues to expand
beyond the core radius.  We shall assume from now on that
$r_{S,0}>r_0$.
 
\begin{figure*}
\includegraphics[width=\textwidth]{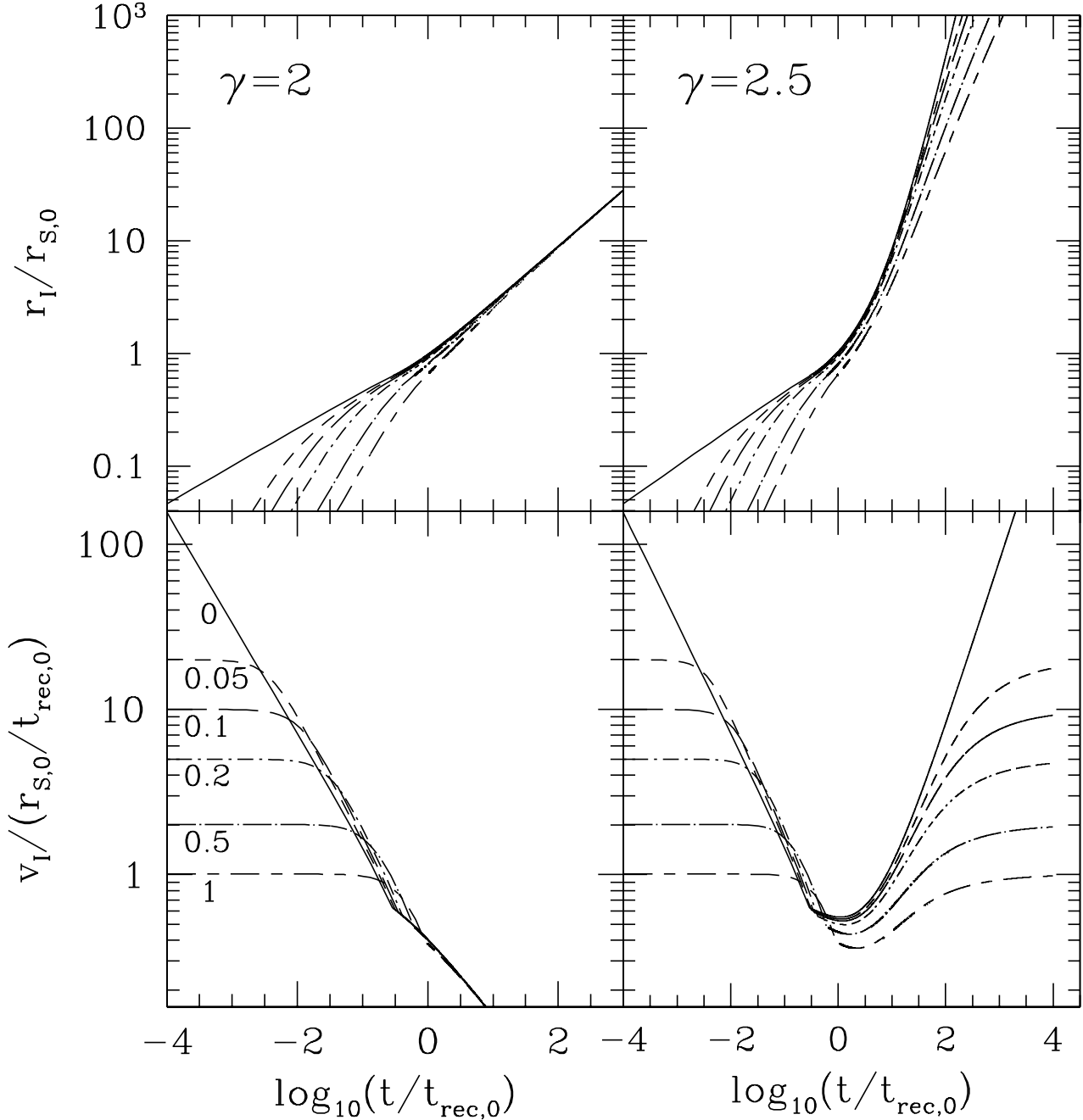}
\caption{ Relativistic I-front for a steady source in a static gas
with a power-law density profile, $n_H\propto r^{-\gamma}$, and a
constant density core for $r\leq r_0$ (left panels for $\gamma=2$,
right panels for $\gamma=2.5$): (a) (top) radius (in units of core
Str\"omgren radius   $r_{\rm S,0}$) and (b) (bottom) velocity (in
units of  $r_{\rm S,0}/t_{\rm rec,0}$).  Curves are labeled by values
of the dimensionless light-crossing time of the core Str\"omgren
radius, expressed in units of the core recombination time, $q\equiv
r_{\rm S,0}/(ct_{\rm rec,0})$.   In these units, the speed of light is
$q^{-1}$. The case $q=0$ corresponds to the nonrelativistic
solution. All curves assume $Y_0=4^{-1/3}$.
\label{powerlaw_fig}
}
\end{figure*}

The time evolution of the  I-front radius is given by inserting the
power-law density profile in  equation (\ref{powerlaw}) into equation
(\ref{I-front}).  We nondimensionalize the resulting velocity equation
by expressing all radii in units of $r_{S,0}$ and time in units of
$t_{\rm rec,0}\equiv (\alpha_Bn_0C)^{-1}$, the  recombination time in
the core 
\be Y\equiv \frac{r}{r_{S,0}}, \ee so $Y_I\equiv
r_I/r_{S,0}$, $Y_0\equiv r_0/r_{S,0}$, and $Y_S\equiv r_S/r_{S,0}$,
while \be w\equiv \frac{t}{t_{\rm rec,0}}.   \ee We also define the
dimensionless ratio of the light crossing and the recombination time
in the core as \be q\equiv \frac{r_{S,0}}{ct_{\rm rec,0}}.  \ee The
solutions for the I-front radius and velocity for each value of the
density profile slope $\gamma$ are then fully characterized by the two
dimensionless parameters $q$ and $Y_0$, as follows:
\begin{equation}
\begin{displaystyle}
\large \frac{dY_I}{dw}=\left\{
\begin{array}{ll}
\frac{1-\left(\frac{2\gamma}{2\gamma-3}\right)Y_0^3
+\left(\frac{3}{2\gamma-3}\right)Y_0^{2\gamma}Y_I^{3-2\gamma}}
{3Y_0^{\gamma}
Y_I^{2-\gamma}+q\left[1-\left(\frac{2\gamma}{2\gamma-3}\right)Y_0^3
+\left(\frac{3}{2\gamma-3}\right)Y_0^{2\gamma}Y_I^{3-2\gamma}\right]},
&\mbox{\normalsize for $\gamma\neq 3/2$},\\ [3mm] 
\frac{Y_0^{-3}-1-3\ln (Y_I/Y_0)}
{\left(\frac{3}{Y_0}\right)\left(\frac{Y_I}{Y_0}\right)^{1/2}+
q\left[Y_0^{-3}-1-3\ln (Y_I/Y_0)\right]},
&\mbox{\normalsize for $\gamma=3/2$}.
\end{array}
\right.
\end{displaystyle}
\end{equation}
For comparison, the speed of light in these dimensionless units  is
$q^{-1}$.

The I-front leaves the core with initial velocity \be
\left.\frac{dY_I}{dw}\right|_{Y_I=Y_0}=\frac{1-Y_0^3}{3Y_0^2+q(1-Y_0^3)},
\label{I-front_init}
\ee which is relativistic (roughly) if $q\ga (3Y_0^2)/(1-Y_0^3)=
[v_{\rm I,NR}(r_0)/(r_{\rm S,0}/t_{\rm rec,0})]^{-1}$, where $v_{\rm
I,NR}(r_0)$ is the nonrelativistic I-front solution speed when
$r_I=r_0$ (i.e. take q=0 in eq. (\ref{I-front_init}) above).  Hence,
for any given $r_0$ and $n_0$, a sufficiently high luminosity
$\dot{N}_\gamma$ is required to make the I-front velocity still
relativistic once the front reaches $r_I=r_0$.  It is possible for the
I-front to accelerate afterwards, however, depending upon the values
of $\gamma$ and $Y_0$, so even if the front is not relativistic when
it leaves the core, it may become relativistic at larger radii.
 
\cite{1990ApJ...349..126F} derived some of the properties of the
R-type I-front phase for the nonrelativistic solution of this
problem. As discussed in \S~\ref{static_subsec}, we can use this
nonrelativistic solution directly to  derive additional properties of
the relativistic solution, as well.  The  nonrelativistic I-front has
a velocity $v_{\rm I,NR}$ which depends upon its radius $r_I$ as
follows: \be v_{\rm I,NR}(r_I)=\frac{v_{\rm
I,NR}(r_0)}{Y_0^{-3}-1}u(\gamma),
\label{vi_NR}
\ee  where $u(\gamma)$ is given by  \ba
u(\gamma)=\left\{\begin{array}{ll}
\left(Y_I/Y_0\right)^{\gamma-2}\left[Y_0^{-3}-
\frac{2\gamma}{2\gamma-3}+\frac{3}{2\gamma-3}\left(Y_I/Y_0\right)^{3-2\gamma}
\right],&\mbox{for $\gamma \neq 3/2$,}\\
\left(Y_I/Y_0\right)^{1/2}\left[Y_0^{-3}-1-3\ln
Y_I/Y_0\right],&\mbox{for  $\gamma=3/2$}.
\end{array}
\right.
\label{complic}
\ea Equations~(\ref{vi_NR}) and (\ref{complic}) then yield the
correct relativistic velocity $v_I(r_I)$ for the I-front if we insert
this $v_{\rm I,NR}$ given above into equation~(\ref{v_I_from_NR}).

For any I-front that expands beyond the radius of the core, the
relativistic I-front will only expand until it reaches the same
Str\"omgren sphere radius $r_S$ as in the nonrelativistic solution,
given by \ba \frac{r_{S}(\gamma)}{r_{\rm S,0}}=\left\{
\begin{array}{ll}
\left[\frac{3-2\gamma}3+\frac{2\gamma}{3}Y_0^{3}
\right]^{\frac{1}{3-2\gamma}} Y_0^{\frac{2\gamma}{2\gamma-3}},
&\mbox{for $\gamma\neq 3/2$},\\ [3mm]  Y_0\exp\left\{\frac13\left[
Y_0^{-3}-1\right]\right\},&\mbox{for $\gamma=3/2$}
\end{array}
\right.
\label{complic2}
\ea \citep{1990ApJ...349..126F}. Finally, there is a (flux-dependent)
critical value of the logarithmic slope of the density profile,
$-\gamma_f$, below which there is no finite Str\"omgren radius $r_S$
at which the I-front is trapped, given by \be
\gamma_f=\frac32\left[1-Y_0^3\right]^{-1}
\label{alpha_f}
\ee \citep{1990ApJ...349..126F}. For density profiles that decline
more steeply than $r^{-\gamma_f}$, the relativistic I-front expands
without bound, just as it does in the nonrelativistic solution.

The relativistic and nonrelativistic I-front propagation solutions for
power-law density profiles are plotted in Figure~\ref{powerlaw_fig}
for the  illustrative cases of $\gamma=2$ and 2.5, for the particular
value of $Y_0=4^{-1/3}$ for which  the critical slope $\gamma_f=2$.
In that case, for $\gamma=2$, the I-front is never trapped at a finite
Str\"omgren radius, but it decelerates continuously and reaches zero
velocity at infinite radius.  Such I-fronts are therefore relativistic
only at early times.  For $\gamma=2.5>\gamma_f$, on the other hand,
the I-front reaches a minimum velocity and thereafter accelerates, so
it is  relativistic both at early and late times.

For example, let us consider a source producing $f_\gamma= 250$
ionizing photons per atom within a lifetime of $t_{s}=3$ Myr, inside a
galactic halo of total mass $10^{11}M_\odot$, switching on at
$z=7$. This corresponds to one of the illustrative cases for the
cosmological H~II regions of individual sources during the
reionization epoch considered in \citet{2005ApJ...624..491I}, except
that here we are interested in the early phase of I-front propagation
{\it inside} the halo, while  \citet{2005ApJ...624..491I} were
interested in its subsequent propagation in the external IGM. For
cosmological halos in the current $\Lambda$CDM universe, the typical
density profile logarithmic slopes outside the central region are
steep, in the range -2.1 to  -2.4 (e.g. Truncated Isothermal Spheres
[TIS], Navarro, Frenk and White [NFW],  or Hernquist profiles). Let
us, therefore, assume $\gamma=2.3$ as a typical value, but, for
illustrative purposes, adopt a core density $n_0$ and a radius  $r_0$
related to the halo virial radius $r_{\rm vir}$ as in the TIS model of
\citet{1999MNRAS.307..203S} and \citet{2001MNRAS.325..468I}, for which
$r_{\rm vir}\approx30r_0$.  For this mass halo at this epoch in
$\Lambda$CDM,  $r_0=0.7$ kpc, $n_0=1.64\,\rm cm^{-3}$, and we obtain
$\dot{N}_\gamma=4.616\times10^{55}\rm s^{-1}$ and  $r_{\rm S,0}=8.1$
kpc, implying $q=0.36$ and $Y_0=0.086$.
The I-front velocities are $r_I=r_0$ and $r_I=r_{\rm vir}$ are then
$v_I(r_0)=0.941c$  and $v_I(r_{\rm vir})=0.978c$, respectively. Hence,
in this example, the  I-front is relativistic inside the galaxy at all
times and is accelerating once it exits the core. Even assuming an
order of magnitude smaller rate of production of ionizing photons, the
velocity is still relativistic, starting at $v_I(r_0)=0.62c$ as the
front exits the core, and accelerating to  $v_I(r_{\rm
vir})=0.82c$. Of course, the actual density profile of gas in such a
halo may be affected by radiative cooling and by departures from
spherical symmetry which are not accounted for in this estimate, so
these numbers are only intended to be illustrative.

\subsection{Plane-stratified Medium}

Given that sources tend to form in regions in which gravitational
instability leads to collapse, the medium may not only be centrally
concentrated but may also depart from spherical
symmetry. Gravitational collapse tends to result in flattened
structures, either by a ``pancake'' instability or by the formation of
a rotationally-supported  disk.  It is therefore instructive to
consider the I-front for a point source in a static, plane-stratified
medium, therefore.  This problem was also considered in the
nonrelativistic limit by \citet{1990ApJ...349..126F}.

The density profile is given by  \be n(z)=n_0{\rm sech}^2(z/z_0),
\label{plane}
\ee where $n_0$ is the central density, $z$ is the height above the
central plane at $z=0$, and $z_0$ is the scale height.  Once again, we
define the dimensionless parameters \be Y_0\equiv\frac{z_0}{r_{\rm
S,0}} \ee
\begin{figure*}
\includegraphics[width=\textwidth]{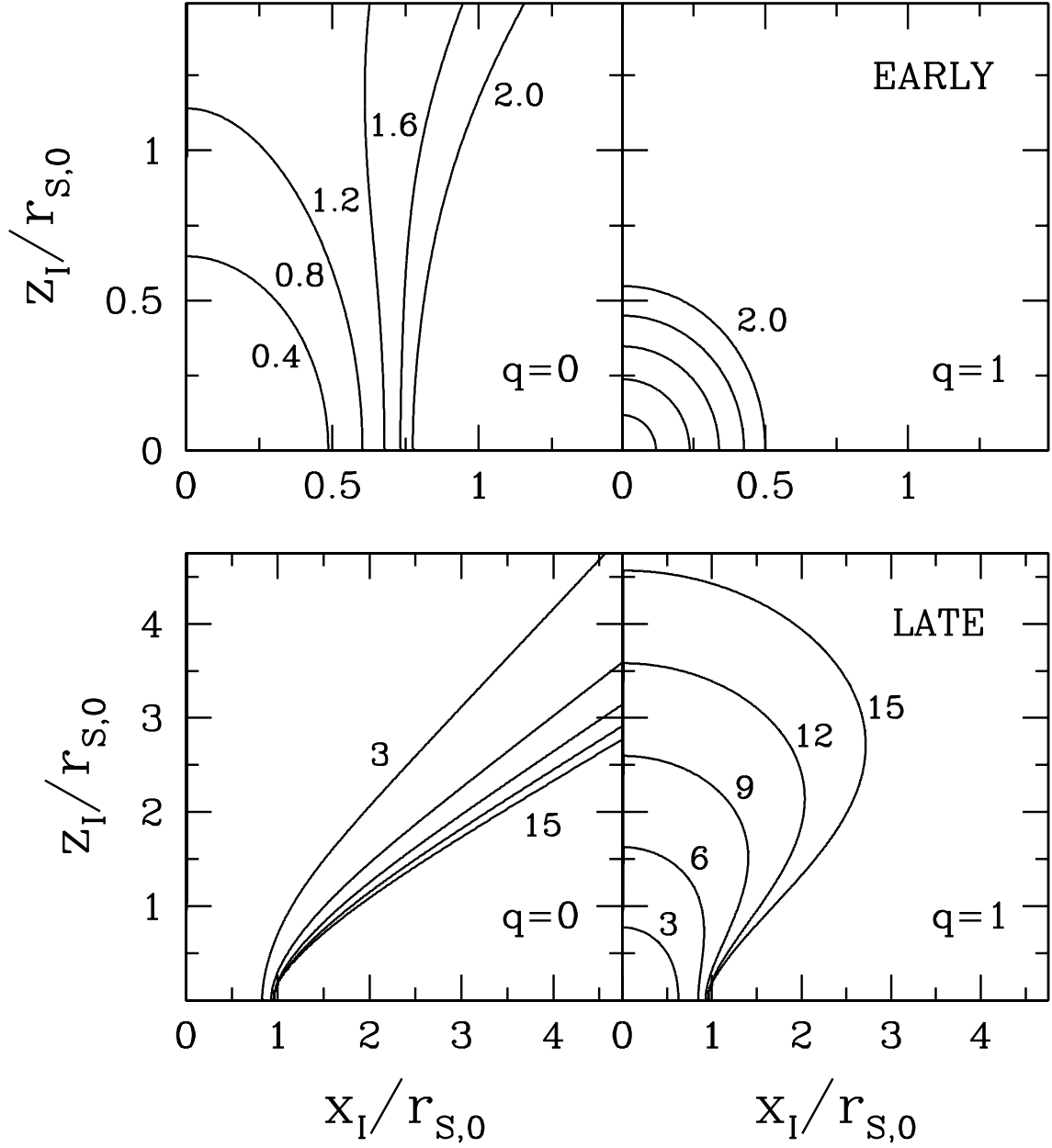}
\caption{Two-dimensional, axisymmetric I-front surfaces  $(x_I/r_{\rm
  S,0},z_I/r_{\rm S,0})$ (in units of the Str\"omgren radius at the
  central density) for a steady source in a static, plane-stratified
  medium at different times after turn-on, $t/t_{\rm rec,0}$ (where
  $t_{\rm rec,0}$ is recombination time at the origin), as labelled,
  for $q=0$ (i.e. nonrelativistic solution) (left panels) and $q=1$
  (relativistic) (right panels).  Both curves assume $Y_0=1/2$.
\label{plane-strat_fig}
}
\end{figure*}
and \be q\equiv\frac{r_{\rm S,0}}{ct_{\rm rec,0}}, \ee where $r_{\rm
S,0}$ and $t_{\rm rec,0}$ are the central Str\"omgren radius and
recombination time, respectively.
\begin{figure*}
\includegraphics[width=\textwidth]{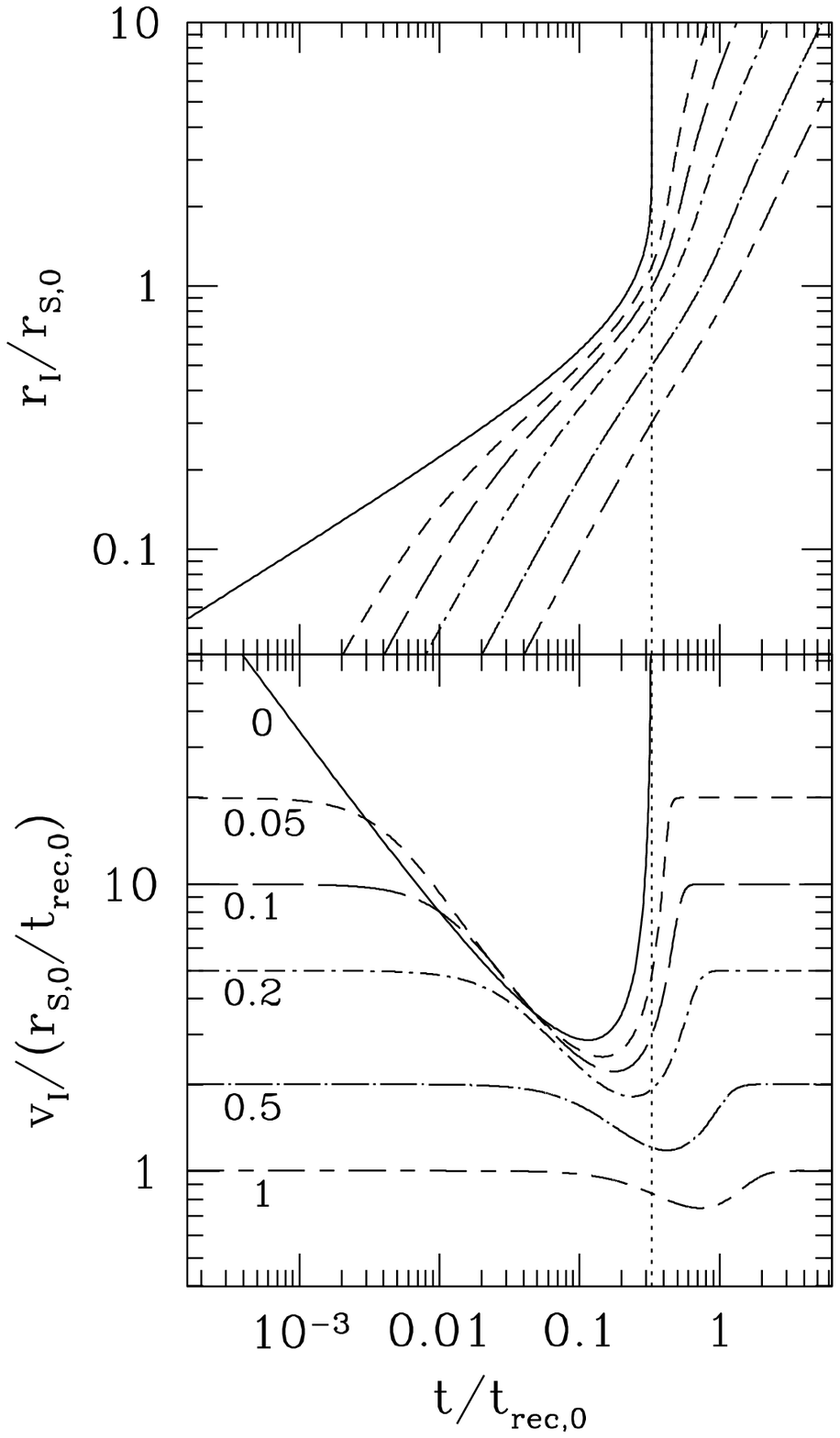}
\caption{ Relativistic I-front radius (upper panel) (in units of
Str\"omgren radius $r_{\rm S,0}$ at central density) and velocity
(lower panel) (in units of $r_{\rm S,0}/t_{\rm rec,0}$) along the
symmetry axis (z-axis) versus time (in units of central recombination
time $t_{\rm rec,0}$) for same plane-stratified case as shown in
Figure~\ref{plane-strat_fig}.  Each curve is for a different value of
$q=r_{\rm S,0}/(ct_{\rm rec,0})$ (as labelled in lower panel), with
$q=0$ corresponding to the nonrelativistic solution (solid).  In these
units, the speed of light is $q^{-1}$.  Vertical dotted line marks the
finite time $t_\infty$ at which the NR solution (q=0) reaches infinite
radius: $z_I(t_\infty)=\infty$.  All curves assume $Y_0=1/2$.
\label{plane-strat_figv}}
\end{figure*}
Application of equation (\ref{I-front}) to the stratified density profile of equation 
(\ref{plane}) leads to the following angle-dependent differential equation for the evolution
of the I-front radius,
\be
\frac{dY_I}{dw}=\frac{1-f(Y_I,\theta)}{3Y_I^2g(Y_I,\theta)+q(1-f(Y_I,\theta))},
\label{planeeq}
\ee
where $Y_I(\theta)\equiv r_I(\theta)/r_{\rm S,0}$, $w\equiv t/t_{\rm rec,0}$,
\be
f(Y,\theta)=\frac{0.66Y_0^3}{\sin^3(\theta)}\tanh^3
\left(\frac{Y\sin(\theta)}{0.88Y_0}\right),
\label{planef}
\ee
and
\be
g(Y,\theta)={\rm sech}^2\left[Y/Y_0\sin(\theta)\right].
\label{planeg}
\ee
In deriving equations (\ref{planeeq})-(\ref{planeg}), we have measured
the polar angle $\theta$ of the direction from the central source to a 
point on the I-front with respect to the $z=0$ plane, so $\theta=\pi/2$
corresponds to the symmetry axis (i.e. the $z$-axis).  We have also
followed Franco et al. (1990) in evaluating the recombination integral
in equation (\ref{I-front}) along each direction by approximating the integral 
using
\be
\int_0^p p^2{\rm sech}^4(p)dp\simeq 0.22{\rm tanh}^3\left(\frac{p}{0.88}\right),
\ee
where $p\equiv (r/z_0)\sin\theta$.

Illustrative solutions of equations (\ref{planeeq})-(\ref{planeg}) for 
relativistic and nonrelativistic I-fronts in a plane-stratified medium
are plotted in Figures \ref{plane-strat_fig} and \ref{plane-strat_figv}.
We adopt the value $Y_0=1/2$ in all cases.  Figure~\ref{plane-strat_fig} 
shows the two-dimensional,
axisymmetric I-front surfaces at different times for the nonrelativistic
solution ($q=0$) and for the relativistic solution for $q=1$.  The NR I-front
solution starts out at superluminal speeds and decelerates in all directions.
Along the symmetry axis, the NR I-front eventually reaches a minimum velocity
and, thereafter, accelerates upward.  This acceleration leads to a superluminal
``blow out'' in which the NR I-front reaches infinite height in a finite time
$t_\infty$.  In the perpendicular direction, however, along the plane of symmetry
at $z=0$, the NR I-front decelerates continuously and approaches the Str\"omgren
radius of a uniform sphere of the same density. The relativistic I-front also 
starts out decelerating but remains close to the speed of light in all
directions, so its shape is initially quite spherical. Like the NR I-front,
the relativistic I-front also reaches a minimum speed along the symmetry axis,
before it accelerates upward once again and approaches the speed of
light. Since its speed is always finite, however, the relativistic I-front
cannot ``blow-out'' as the NR front does in a finite time. Since, in the
central plane, the relativistic I-front slows to approach the same Str\"omgren
radius as does the NR solution, above this plane it must balloon upward and
outward, confined at the waist by this ``Str\"omgren belt''. 

\section{Cosmological Relativistic I-fronts} 
\label{cosmo_sect}

Next we generalize the treatment by \citet{1987ApJ...321L.107S} of an I-front
in a uniformly expanding cosmological background at a redshift $z$, to take
account of special relativity.  Here it is convenient to define the 
cosmological scale factor as $a=(1+z_i)/(1+z)$, where $z_i$ is the redshift of
the source turn-on, corresponding to initial time $t_i$. 
The comoving hydrogen  number density ($n_{H,i}$), 
I-front radius ($r_I$), and velocity  ($v_{I}$), then correspond
to quantities in proper coordinates (in the frame of the source) 
as $n_{H,p}=n_{H,i}/a^3$, $r_{I,p}=ar_I$, and
$v_{I,p}=dr_{I,p}/dt=d(ar_I)/dt,$ which means that these comoving
quantities are defined {\em at the source turn-on time}, not at the present
as is often done. 

\subsection{Relativistic I-front Velocity in the Cosmologically Expanding IGM}

For a relativistic I-front in the cosmologically expanding IGM at redshift
$z$, the velocities of the front and of the gas just outside it as measured
in the lab frame (i.e. the source rest frame) are the proper velocities, 
$v_{I,p}=\beta_{I,p}c$ and $v_{g,1,p}=\beta_{g,1,p}c$, respectively.  This 
latter gas proper velocity is the sum of the proper Hubble velocity at the
location of the front at the proper radius $r_{I,p}$, $v_h(r_{I,p})=
\beta_h(r_{I,p})c$, and its proper peculiar velocity, $v_{g,pec}$,
\ba
v_{g,1,p}=v_h(r_{I,p})+v_{g,pec}\nonumber\\
=H(z)r_{I,p}+v_{g,pec},
\label{app2}
\ea
where $H(z)$ is  the Hubble constant at redshift $z$. We are interested in 
I-front radii that are small compared with the horizon,
so $Hr_{I,p}\ll c$, and in situations in which $v_{g,pec} \ll c$ as well.
For the problem of an I-front in the uniformly-expanding, mean IGM, in fact,
we can set $v_{g,pec}=0$.  The I-front velocity measured by the gas at that
location is the proper peculiar velocity $v_{I,pec}'$, given by equation 
(\ref{app1}) as 
\be
v_{I,pec}'=
\frac{v_{I,p}-v_h(r_{I,p})}{1-\frac{v_h(r_{I,p})v_{I,p}}{c^2}}
=\frac{v_{I,p}-H(z)r_{I,p}}{1-\frac{H(z)r_{I,p}v_{I,p}}{c^2}}.
\label{app3}
\ee
Let us define the proper peculiar velocity of the I-front as measured in the
lab frame of the source as $v_{I,pec}=\beta_{I,pec}c$, where
\be
v_{I,pec}=v_{I,p}-H(z)r_{I,p}=a\frac{dr_I}{dt}.
\label{app4}
\ee
Equations (\ref{app3}) and (\ref{app4}) then yield
\be
v_{I,pec}'=\beta_{I,pec}'c=\frac{\beta_{I,pec}c}{1-\beta_h\beta_{I,p}}.
\label{app5}
\ee

According to equation~(\ref{app10}), the proper peculiar velocity of the
I-front in the lab frame of the source is given in terms of these quantities by
\be
\beta_{\rm I,pec}=
\left[\frac{(1-\beta_{I,p}\beta_{h})}{\gamma_{\rm I,pec}'}\right]
\left[\frac{\gamma_{I,p}(1-\beta_{I,p})F}{\beta_in_{H,p}c}\right],
\label{beta_pec_cosmo}
\ee
where $\beta_{I,p}=\beta_{\rm I,pec}+\beta_h$, $\gamma'_{\rm
  I,pec}\equiv\gamma(\beta'_{\rm I,pec})$, and we have replaced $n_{H,1}$ by
$n_{H,p}\gamma_h$ (using equation~[\ref{app11}]). Substituting
equation~(\ref{app5}) into equation~(\ref{beta_pec_cosmo}) yields an equation
that can be solved for $\beta_{\rm I,pec}$ as a function of $\beta_h$ and the
quantity $F/(\beta_in_{H,p}c)$. There are two roots to this equation,
corresponding to $\beta_{\rm I,pec}>1$ and $\beta_{\rm I,pec}<1$,
respectively. The physical solution must have $\beta_{\rm I,pec}<1$, which
yields 
\be
\beta_{\rm I,pec}=\frac{a}{c}\frac{dr_I}{dt}
=\frac{(1-\beta_h)F}{\beta_in_{H,p}\gamma_hc+F}.
\label{betaIpec}
\ee 
The photon flux $F$ in equation~(\ref{betaIpec}) is that at the location of
the I-front as measured in the frame of the source, given by
equations~(\ref{app21}) and (\ref{flux_at_front}) as 
\be
F(r_I,t)=\frac{S(r_I,t_R)}{4\pi a^2r_I^2},
\label{flux_at_front_cosmo}
\ee
with 
\be
S(r_I,t_{R,I})=\dot{N}_\gamma(t_{R,I})-\int_0^{ar_I}dr_p\cdot 4\pi
r_p^2\chi_{\rm eff}x^2n_{H,p}^2\gamma_h^{-2} \alpha_BC,
\label{rec_corr_cosmo}
\ee
where the proper radius $r_p$ is related to the comoving radius $r$ by
$r_p=ar$, the retarded time is $t_{\rm R,I}=t-r_{\rm I,p}/c$ and the
quantities inside the integral are evaluated at the time 
$t'=t_{\rm R,I}+r_p/c$.

The Hubble velocities across the H~II regions of interest to us here are
always much smaller than the speed of light. For a flat universe in the
matter-dominated era, with total matter density $\Omega_0$ in units of the
critical density and Hubble constant $h$ today in units of
$100\rm\,km\,s^{-1}Mpc^{-1}$, cosmic expansion results in a Hubble velocity
at the proper radius, $r_{\rm I,p}$, of the I-front, given by
\be
\beta_h(r_{\rm I,p},z)=\frac{H(z)r_{\rm I,p}}{c}\approx4.54\times10^{-3}
\left(\frac{\Omega_0h^2}{0.15}\right)^{1/2} \left(\frac{1+z}{7}\right)^{1/2}
\left(\frac{r_{I,p}}{6\,{\rm Mpc}}\right).
\label{hubble_small}
\ee 
Since the largest H~II regions (e.g. around rare luminous QSOs) are estimated
to be several Mpc in proper radius at $z\sim6$, this implies that $\beta_h$ is
always less than of order 1\%. In that case, we can simplify
equations~(\ref{betaIpec})-(\ref{rec_corr_cosmo}) above by setting $\beta_h=0$
and $\gamma_h=1$, to yield
\be
\beta_{\rm I,pec}=\frac{a}c\frac{dr_I}{dt}=\frac{F}{\beta_in_{H,p}c+F},
\label{betaIpec_simple}
\ee  
where
\be
F=\frac{\dot{N}_{\gamma}(t_R)-{4\pi}r_I^3a^{-3}{n_{H,i}}^2
       C\alpha_B\chi_{\rm eff}/3}{4\pi a^2r_{I}^2}. 
\label{f_cosmo}
\ee
In terms of the comoving volume , $V_I=4\pi r_I^3/3$,
equations~(\ref{betaIpec_simple}) and (\ref{f_cosmo}) yield
\be
\frac{dV_I}{dt} 
=\frac{\dot N_\gamma(t_R)-C\alpha_B\chi_{\rm eff}n_{H,i}^2a^{-3}V_I}
{\beta_in_{H,i}+\frac{a}{4\pi r_I^2c}
\left[\dot N_\gamma(t_R)-C\alpha_B\chi_{\rm eff}n_{H,i}^2a^{-3}V_I\right]}.
\label{I-front_cosmo}
\ee 
Equation~(\ref{I-front_cosmo}) can also be derived from the
cosmological equivalent of the photon conservation
equation~(\ref{eq:global_conserv}).  This last approach was first proposed 
(in a simplified form that did not account for recombinations 
or the presence of helium) by \citet{2003AJ.126..1W} (in their Appendix).
Equation~(\ref{I-front_cosmo}) was previously derived, including the effect of
recombinations, but not of helium, by \citet{2005ApJ...620...31Y} using 
a different approach from ours.

Following the approach in \citet{1987ApJ...321L.107S}, we define the
dimensionless volume as 
\be
y\equiv V_I/V_{S,i},
\ee
where $V_{S,i}$ is the {\em initial} Str\"omgren volume,
$V_{S,i} = 4\pi r_{S,i}^3/3
=\dot{N}_{\gamma,i} (\alpha_B C\chi_{\rm eff}n^2_{H,i})^{-1}$,
and the dimensionless time as
\be
x \equiv t/t_i,
\ee
which is scaled in terms of the initial time rather than in terms of the
recombination time as $w$ was above. Substituting these relations into 
equation~(\ref{I-front_cosmo}), and using the fact that, for the
matter-dominated era, $x=a^{3/2}$, we obtain (for $\beta_i=\chi_{\rm eff}$)
\be
\frac{dy}{dx}=\frac{\lambda[l(x_R)-y/x^2]}
{1+(q \chi_{\rm eff}^{-1}/3)(x/y)^{2/3}[l(x_R)-y/x^2]}, 
\label{nondim_cosmo0}
\ee
where
\be
\lambda\equiv \frac{t_i}{\chi_{\rm eff}t_{\rm rec,i}}=
t_i C\alpha_B n_{H,i},
\ee
is the ratio of the age of the universe to the recombination time when the 
source turned on, divided by $\chi_{\rm eff}$,
\be
q\equiv \frac{r_{S,i}}{c t_{\rm rec,i}}
= \frac{r_{S,i} \chi_{\rm eff} \lambda}{c t_i},
\ee
is the light-crossing time of the initial Str\"omgren radius, in units of the
initial recombination time, and 
\be
l(x)\equiv \dot{N}_\gamma(t)/\dot{N}_\gamma(t_i)\equiv
\dot{N}_\gamma(x)/\dot{N}_{\gamma,i}
\ee
is the ionizing photon luminosity in units of its initial value
(i.e. $l(x)\equiv 1$ for a steady source). For $q=0$ ($c\rightarrow\infty$),
equation~(\ref{nondim_cosmo0}) reduces to the nonrelativistic form derived by 
\citet{1987ApJ...321L.107S}, for which they gave an exact analytical
solution for the case of a steady source, as follows:
\be
y_{\rm NR}(x)={\lambda}e^{\lambda/x}
\left[x {\rm Ei}(2,{\lambda}/{x})-{\rm Ei}(2,\lambda)\right],
\label{y_exact}
\ee 
where ${\rm Ei}(2,x)\equiv\int_1^\infty \frac{e^{-xt}}{t^2}dt$ 
is the Exponential
integral of second order. This exact solution also
applies to a flat universe with a cosmological constant in the matter-dominated
era \citep{2005ApJ...624..491I}. As described in \S~\ref{static_subsec}, we
can use this exact solution of the NR I-front equations to derive the exact
solution of the {\it relativistic} I-front equations (i.e. arbitrary $q\neq0$)
directly, if we replace the time $t$ in the NR solution by the retarded time
$t_R$. In that case, the exact, analytical, relativistic I-front solution is
given by the following implicit equations:
\be
y(x)=y_{\rm NR}(x_R)={\lambda}e^{\lambda/x_R}
\left[x_R {\rm Ei}(2,{\lambda}/{x_R})-{\rm Ei}(2,\lambda)\right],
\label{y_exact_rel}
\ee
and
\be
x=x_R+x^{2/3}y_{\rm NR}^{1/3}(x_R)\left(\frac{q}{\lambda\chi_{\rm eff}}\right),
\label{x_exact_rel}
\ee
where we have used the fact that $a(x)=x^{2/3}$, $x=x_R+r_{I,p}(x)/(ct_i)$,
and $r_{I,p}(x)=a(x)r_{S,i}[y_{\rm NR}(x_R)]^{1/3}$. Equations~(\ref{y_exact_rel})
and (\ref{x_exact_rel}) easily reduce to a cubic equation for $x(x_R)$, which
has an exact analytical solution with one physical root. Since it is a rather
lengthy expression, we omit it here. 

\begin{figure*}
\includegraphics[width=\textwidth]{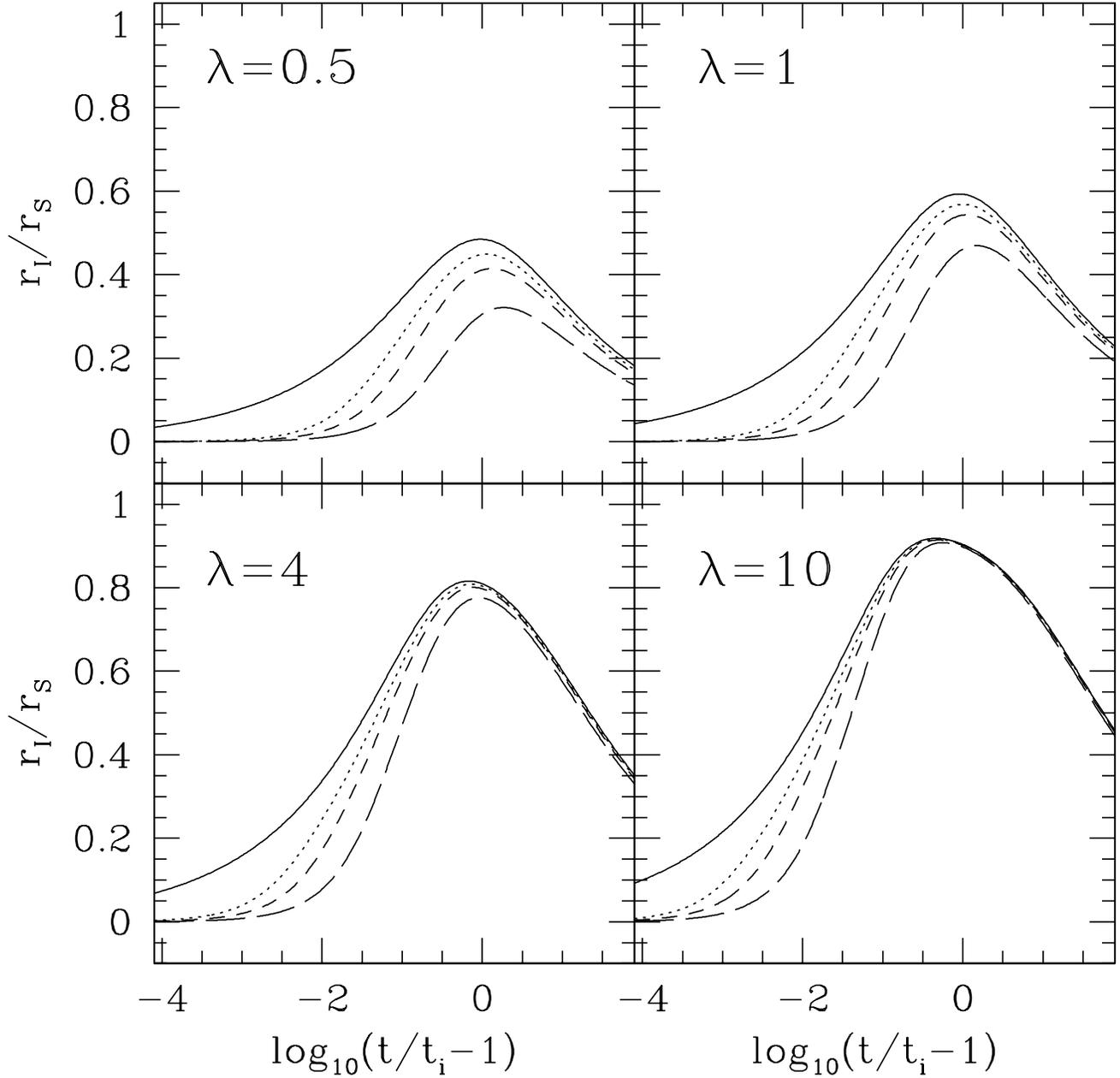}
\caption{
Cosmological I-front radius versus time for steady source in the mean IGM, with
comoving radius in units of the time-varying, comoving Str\"omgren radius,
$r_S(t)=a(t)r_S(t_i)$, and time in units of the age of universe $t_i$ at
source turn-on.  Different curves in each panel correspond to different values
of $q=r_{S,i}/(ct_{rec,i})$, with $q=0$ 
(i.e. nonrelativistic limit, top line) and (from top to bottom) 
$q=0.1,0.2$ and 0.5.  Each panel is for different values of 
$\lambda\equiv{t_i}/{(\chi_{\rm eff}t_{\rm rec,i})}$, as 
labeled.  We have assumed that $\chi_{\rm eff}=1$ for simplicity.
\label{nondim_soln_figc1}
}
\end{figure*}

\begin{figure*}
\includegraphics[width=\textwidth]{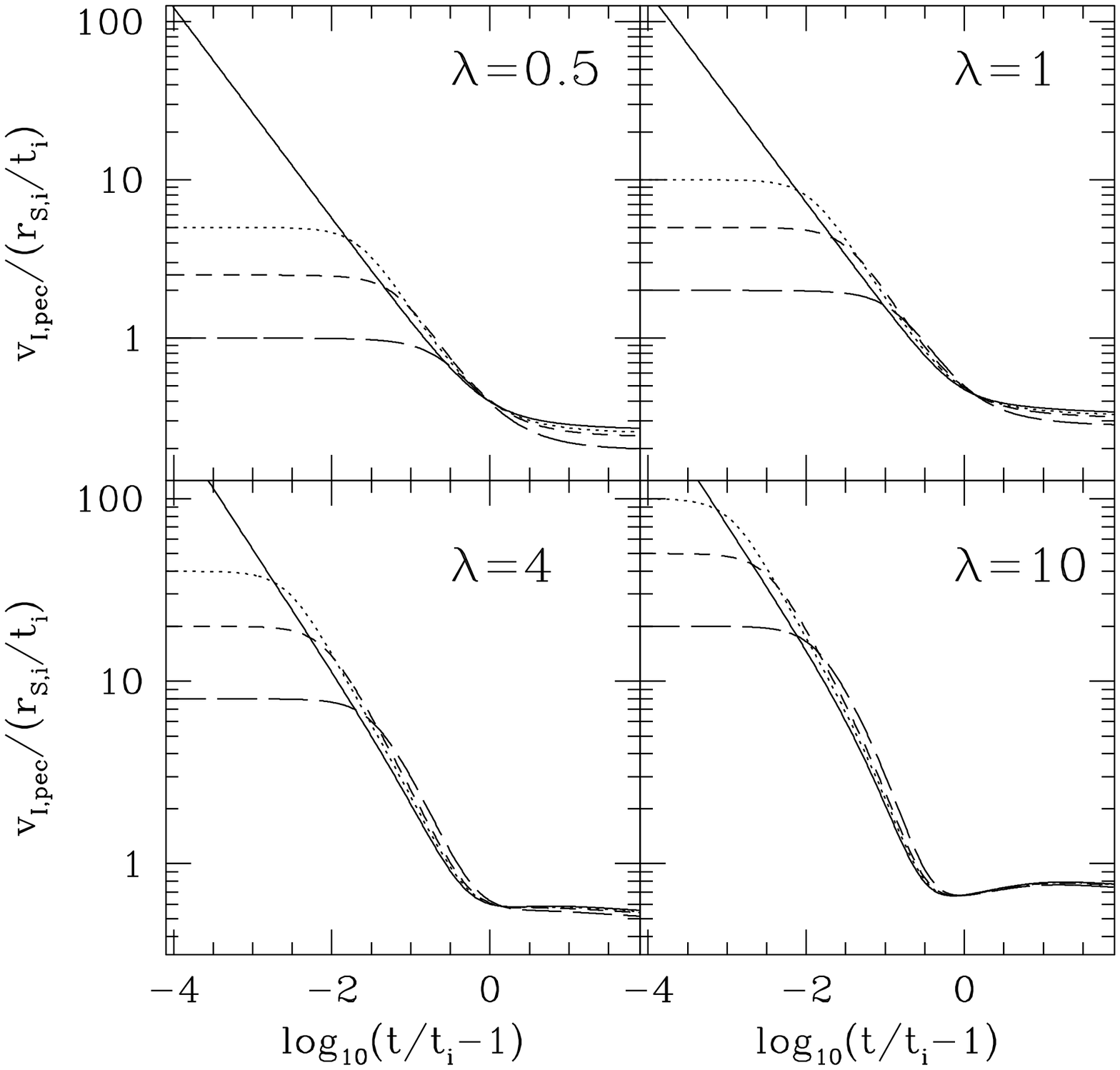}
\caption{
Same as Figure~\ref{nondim_soln_figc1} but for the I-front peculiar velocity,
instead, where $v_{\rm I,pec}$ is in units of $(r_{S,i}/t_i)$.  The speed of light,
$c$, corresponds in these units to $(\chi_{eff}\lambda/q)$.  We assume 
$\chi_{eff}=1$ for simplicity.
\label{nondim_soln_figc2}
}
\end{figure*}

For each value of $\lambda,$ we obtain a family of solutions for different
values of $q$. In fiducial units 
\ba
\lambda&=&0.822C{(1+z_i)_{10}^{3/2}}
         \left(\frac{\Omega_0h^2}{0.15}\right)^{-1/2}
\left(\frac{\Omega_bh^2}{0.022}\right)
\left(\frac{\chi_{\rm eff}}{1.08}\right)\\            
q&=&0.0217\chi^{2/3}_{\rm eff}C^{2/3}\dot{N}_{\gamma,56}^{1/3}(1+z_i)_{10}^{-1/2}
\left(\frac{\Omega_bh^2}{0.022}\right)^{1/3},
\label{params}
\ea
where $(1+z_i)_{10}\equiv(1+z_i)/10,$
$\dot{N}_{\gamma,56}\equiv\dot{N}_{\gamma}/(10^{56}\rm \,s^{-1})$, and
$\Omega_b$ is the mean baryonic density in units of the critical density. 
The solutions for  
$\lambda=0.5,1,4,$ and 10 and $q\chi_{\rm eff}^{-2/3}=0,0.1,0.2$ and 0.5 
are shown in Figures~\ref{nondim_soln_figc1} and~\ref{nondim_soln_figc2}.
The relativistic 
correction is most significant for small $\lambda$ and large $q$ values,
i.e. at later times, stronger sources and little or no clumping of the
gas. 
This leads to somewhat counterintuitive behavior.
For example, increasing the clumping factor increases $q$, and naively, 
based on equation~(\ref{nondim_cosmo0}),
could be expected to cause a stronger relativistic correction.  However,
it also increases the value of $\lambda$, which results in a smaller 
relativistic correction.  This is because larger clumping of the gas 
increases the recombinations in the H~II region, which decrease
the ionizing flux at the front, slowing it down into the nonrelativistic 
regime.  Below we shall see that infall around the source has a similar effect,
increasing the density near the source and rendering the relativistic 
corrections somewhat  weaker than a simple, uniform-density treatment would 
predict.  

As in the case of a spherical I-front in a static medium
discussed in \S3, the apparent size of a cosmological H~II region
observed at a given time will depend upon the angle between the line
of sight to the source and the line connecting the source to the point
we observe on the I-front surface.  This difference results from the
difference in the light travel times for light received by the
observer from different locations on the front.  If we were to measure
the radius of the H~II region transverse to the line of sight at the
distance of the source, for
example, and interpret this as the proper radius given by the
relativistic solution at some time $t$, then the line-of-sight radius
on the near-side of the source would be given by the {\em nonrelativistic}
solution at the same physical time.  Hence, the line-of-sight radius
will appear to be larger than the transverse radius and will appear to
have expanded toward the observer at superluminal speeds.  This
deviation of the shape of the cosmological H~II region from spherical
was also discussed by Yu (2005) in the context of the H~II regions
around QSO's at redshifts $z>6$ whose spectra show Gunn-Peterson troughs.

\section{Cosmological Relativistic I-fronts in Presence of Small-scale 
Structure and Infall}

\label{MH_sect}

\begin{figure*}
\includegraphics[width=0.45\textwidth]{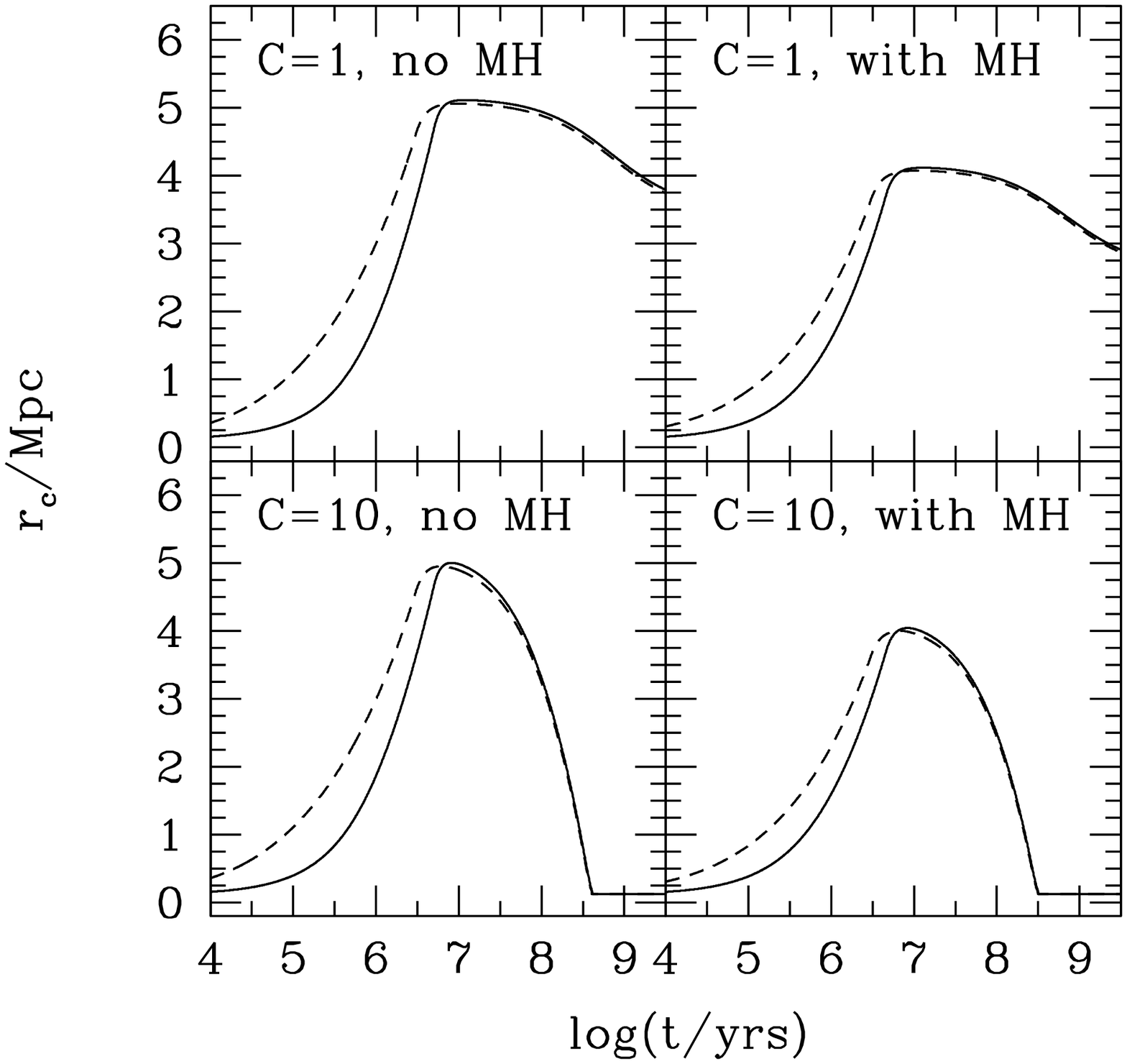}
\includegraphics[width=0.45\textwidth]{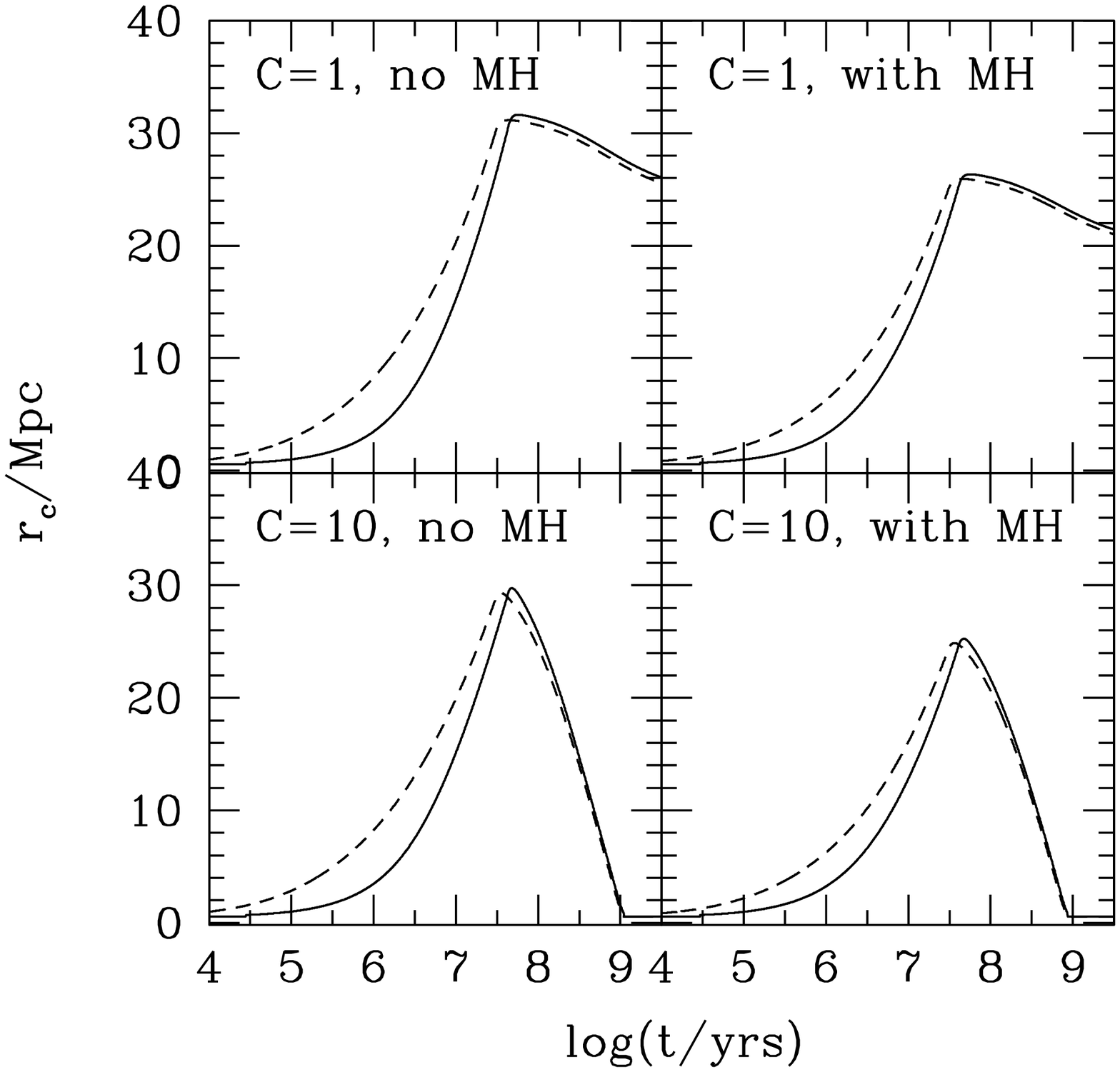}
\caption{Cosmological I-front radius in present ($z=0$) comoving units,
  $r_c=r_I(1+z_i)$, versus time (since 
source turn-on) for a large starburst (left panels) and a luminous QSO (right 
panels) in the local IGM with infall density profile, both starting at
$z_i=7$, with assumed parameters as described in the text, for the cases
with mean ($C=1$) and clumped ($C=10$) IGM, with and without minihalos
present, as labeled, for the relativistic case (solid) and the
corresponding nonrelativistic one (dashed). 
\label{single_fig}}
\end{figure*}

When self-shielded neutral structures (minihalos) are present on the 
path of the
propagating cosmological I-front, we follow the approach in
\citet{2005ApJ...624..491I}. In this case the  I-front jump 
condition in equation~(\ref{betaIpec}) and the I-front evolution
equation~(\ref{I-front_cosmo}) are modified by setting  
\be
 \beta_i \equiv (1-f_{\rm coll})\chi_{\rm eff}+
[1+A({\rm He})]f_{\rm coll,MH}\bar\xi,
\label{eq:betai}
\ee
where $f_{\rm coll}$ is the total collapsed baryon fraction, $f_{\rm coll,
  MH}$ is the collapsed fraction of just the minihalos, and  $\xi$ is 
the number of ionizing photons consumed per minihalo
atom such that $\bar \xi$ is the appropriate average over the
distribution of minihalos at the instantaneous location of the global
I-front. This photon consumption factor is flux-dependent and in the
relativistic case the appropriate flux is $F$, the flux of ionizing
photons in the frame of the source, as the relative velocity between
the source and the minihalos is much smaller than $c$.

The average number of ionizing photons absorbed per 
minihalo atom at the current location  of the I-front, 
 $\bar{\xi}$, is given by integrating the photon
consumption rate $\xi$ per individual minihalo over the Press-Schechter mass
function of minihalos at that epoch. The photon consumption rates for
individual minihalos were determined by \citet{2005MNRAS...361..405I}
using high-resolution, numerical gas dynamics simulations that included
radiative transfer. They have summarized their detailed results by providing
fitting formulae for the individual minihalo consumption rates and their
dependence on halo mass, redshift, incident flux and ionizing source
spectrum. According to \citet{2005MNRAS...361..405I}, 
\be
\xi(M,z,F_0)\equiv 1+4.4M_7^{0.334+0.023\log_{10} M_7}
F_0^{0.199 -0.042 {\rm log_{10}} F_0}(1+z)_{10},
\ee 
for a $5\times 10^4K$ blackbody external ionizing spectrum,  representing
Pop.~II stars (and similar numbers for a QSO spectrum), for a minihalo of 
mass $M_7$ (in units of $10^7 M_\odot$), overtaken by an intergalactic I-front 
at a redshift of $z$ which is driven by an external source of flux $F_0$, the 
flux in units of that from a source emitting 
$10^{56}$ ionizing photons per second at a proper 
distance $d$ of 1 Mpc, i.e.  $F_0 \equiv {\dot{N}_{\gamma,56}}/{d^2_{\rm Mpc}}$.
Again, the relevant flux to use in calculating $\xi$ is $F$, the photon flux 
in the source frame, as it is only the ionization front, not the self-shielded 
structures, that is moving relativistically with respect to the source.
Our model takes account of the biasing of minihalos with respect to the
ionizing sources (i.e. the clustering of minihalos in space) by modifying the
mass function of the minihalos as a function of their distance from the
central source halo in each H~II region. It also includes a simple treatment
of the effect of local infall on the density of the IGM and the clustering of
minihalos surrounding each source [see \citet{2005ApJ...624..491I}
for details]. 

In the results below we specialize to  a Cold Dark Matter (CDM)
cosmological model with parameters $h=0.7$, $\Omega_0$ = 0.3,
$\Omega_\Lambda$ = 0.7, and $\Omega_b = 0.045$,  $\sigma_8 = 0.87$, where
$\Omega_\Lambda$ is  the total vacuum density in units of the critical
density, $\sigma_8^2$ is the variance of linear fluctuations  extrapolated to
the present and filtered on the $8 h^{-1}{\rm Mpc}$ scale, and $n_p$ is the
index of the power spectrum of primordial density fluctuations
\citep{2003ApJS..148..175S}. Note, however, that as the
fronts we are considering are all at high redshifts, the value of
$\Omega_\Lambda$ has no  direct impact on our calculations.  Finally,
the \citet{1999ApJ...511....5E} transfer  function was used.
In Figure~\ref{single_fig}, we show our results for the I-front radius
evolution for two representative sources. The first is a galaxy of 
total mass $10^{11}M_\odot$ at an initial redshift of $z_i=7$ 
(corresponding to a 3-$\sigma$ density fluctuation), which hosts a starburst
that produces a total of $f_\gamma=250$ ionizing photons per baryon within a
lifetime of $t_s=3$ Myr, corresponding to $\dot{N}_\gamma=4.08\times10^{55}\rm
s^{-1}$. For $t>t_s$, the ionizing photon production  
is assumed to drop as $t^{-4.5}$ \citep{2003ApJ...595....1H}.
The second source roughly corresponds to a high-redshift luminous QSO, 
similar to those found by the Sloan Digital Sky Survey 
\citep[e.g.][]{2004AJ....128..515F}. Here we assume an ionizing photon
production rate of $\dot{N}_\gamma=10^{57}\rm s^{-1}$, a source lifetime 
$t_s=30$ Myr starting at $z=7$, and a host halo of $10^{13}M_\odot$. 
For $t>t_s$, we model the drop in the ionizing photon production rate
as an even steeper power of time, $\dot{N}_\gamma(t>t_s)\propto t^{-10},$
although the precise form of this fall-off is
unimportant since we are interested in the expansion 
phase of the H~II region. \footnote{Our I-front propagation equation remains
valid even if the source luminosity decays, as long as the I-front
speed remains positive and the luminosity decay time is greater than
the ionization equilibration time inside the H~II region.  This
equilibration time is much smaller than the fully-ionized
recombination time, however, since it is given by the product of that
recombination time and the small neutral fraction inside the H~II region.
The propagation equation is also valid even when it yields a negative
I-front velocity, but only if the luminosity decay time exceeds the
fully-ionized recombination time.  In that case, the I-front shrinks
to match the instantaneous Str\"omgren sphere.}  We show the
relativistic results (solid) and the na\"ive nonrelativistic results
(dashed) for direct comparison, and consider four cases with mean
($C=1$) and clumped ($C=10$) IGM, with and without minihalos present.  

For both sources, the I-front expansion is initially superluminal in the
nonrelativistic model, causing the H~II region radius to be significantly
larger than the correct, relativistic model.  These differences
diminish somewhat at later times, but remain significant throughout
the expansion phase of the H~II region. For both sources, the IGM clumping has
only a modest effect, since even for $C=10$ the recombination times are longer
than the source lifetimes. The presence of minihalos, on the other hand,
decreases the size of the H~II regions more considerably, by about 25\%
in each case. We have also calculated the I-front expansion for a 
weaker source at higher redshift (a galaxy halo of  total mass
$10^{8}M_\odot$ at initial redshift $z_i=15$, corresponding to  a
$3-\sigma$ density fluctuation, which hosts a Pop. II starburst that
produces a total of $f_\gamma=250$ ionizing photons per baryon within
a lifetime of $t_s=3$ Myr) and, as expected from Figure~\ref{nondim_soln_fig}, 
the relativistic correction in this case was negligible.

Finally, we quantified the impact of relativistic effects  on the global
progress of reionization by recalculating the reionization models described
in detail in \citet{2005ApJ...624..491I}.
For all models, the relativistic corrections were small, decreasing the
global ionized mass fraction at any given time by $\sim 1\%$ as compared 
to the nonrelativistic treatment.  The corresponding effect on the mean 
electron-scattering optical depth $\tau$ was even smaller,
never exceeding a small fraction of 1\%. Thus while the 
relativistic expansion of an I-front about an individual strong source 
close to overlap is important, it can generally be ignored when computing 
the global reionization  history, especially 
given the other significant uncertainties involved.

\section{Summary and Conclusions}
\label{summary_sect}

We have considered the effect of special relativity on the propagation
of ionization fronts. The standard (nonrelativistic) treatment of
I-fronts neglects the finite speed of light and the relative motion of
the radiation source and the gas it ionizes. As a result, when the
ionizing photon-to-atom ratio exceeds unity at the location of the
I-front, unphysical superluminal I-front motion is predicted to occur. We have
generalized the I-front  continuity jump condition and  the equations
that describe the transfer of ionizing radiation 
between the source and  the I-front to take account of these things in
a relativistically-correct way.  By combining these equations, we have derived
the relativistic equation of motion for the R-type I-front created
when a point source of ionizing radiation turns on in a neutral gas,
including the 
possibility that the gas is non-uniform and in motion.  We then
applied this relativistic equation to several
illustrative problems of interest in the theory of H~II regions in the
interstellar and intergalactic media, including the cosmological
problem of the reionization of the universe.  These include the cases of a steady source
in a static gas that is either uniform, spherically-symmetric with a
power-law density profile or plane-stratified.  We have also solved
the problem of a point source in a cosmologically-expanding IGM.   
We have shown that, in all cases, the radius of the
relativistic I-front at a given physical time is identical to that in the
{\it nonrelativistic} solution in which the physical time $t$ is
replaced by the {\it retarded} time, equal to $t$ minus the
light-travel time from the source to the front.

In the case of a static medium, we obtain an exact analytical solution
for the problem of a steady source in a uniform gas, which depends
only on the parameter $q$, the time for light to cross the Str\"
omgren radius, in units of the recombination time. Applying this
solution to the problem of an O-type star or OB association in a
molecular cloud, we find $q$ values of order unity  over the full
range of appropriate densities and fluxes.  However, while this
implies that corrections are formally important, they only affect
ionization fronts during the rapidly-expanding R-type phase, which
lasts only until the I-front expands to fill the Str\"omgren sphere,
and not during the slower D-type hydrodynamic expansion  that
dominates the bulk of the evolution of these H~II regions.

For a source in a {\it nonuniform}, static gas, however, I-fronts can,
in principle, accelerate back up to relativistic speeds if they
propagate down a steep enough density gradient. We have solved this
problem for the cases of a power-law density profile in spherical
symmetry and a plane-stratified gas. These solutions depend both upon
the value of the parameter $q$ calculated at the central density in
the core of these nonuniform profiles and on the dimensionless ratio
of the core radius of the density profile (or the scale-height in the
plane-stratified case) to the Str\"omgren radius.  Such solutions are
relevant, for example, to the ``champagne phase'' of the H~II region,
which results when the D-type I-front from a stellar source in a dense
environment detaches from the shock that leads it during  the
dynamical expansion phase, and the resulting R-type front runs away
towards large radii. In that case, relativistic corrections are
important, not only right after the source turns on, but
at late times as well.

In the cosmological case, we have derived the exact, analytical solution for
the dependence of the relativistic I-front radius on time, by generalizing the
solution derived by \citet{1987ApJ...321L.107S}. We find a complete
family of solutions that are described by two dimensionless parameters: $q$,
which is now defined as the light crossing time at  the
Str\" omgren radius in units of the {\em initial} recombination  time,
and $\lambda$, the age of the universe at source turn-on in units of
the initial recombination time.  In this case, relativistic effects are
again important only if $q \gtrsim 0.1$ , but now only if 
$\lambda \lesssim 1$ as well.  Thus relativistic corrections
are significant only in cases in which {\em both} the source flux
is large (corresponding to a large $q$) and the redshift is 
relatively low (corresponding to a small $\lambda$).

Incorporating relativistic effects into the more exact modeling
described in \citet{2005ApJ...624..491I} allows us to
assess the importance of relativistic effects in a
realistic reionization scenario that includes both infall and the
impact of small-scale (self-shielding) structures.  In the individual
case of a large galaxy or luminous QSO at the end of the reionization
epoch ($z \sim 7$), relativistic effects can be quite significant:
resulting in $\approx 4$ and 2 times smaller ionized volumes after
1 and 3 Myrs, respectively, in the galactic case, and
$\approx 20$ and 2 times smaller volumes after 1 and 10 Myrs, 
respectively, in the QSO case.  The relative differences then decrease with 
time, but remain significant until the source luminosity decays.
Afterwards, at time 
$t\sim2t_{s}$, the relativistic solution for the H~II  region radius
finally matches the nonrelativistic one. Although the I-front
propagation becomes subluminal earlier than this, and thus the usual
nonrelativistic expression for its speed becomes correct, the radius
of the H~II region remains smaller throughout the
source lifetime. Furthermore, the  presence of small-scale structure
can decrease the size of the H~II region  by  up to an additional
$\sim 25$\%.

Generalizing these results to the issue of global reionization, however,
yields far less dramatic changes. As most of the intergalactic medium was 
ionized  by smaller (lower $q$)  and higher-redshift (higher $\lambda$) 
sources, we find that relativistic corrections change the 
ionized mass fraction at any give time by $\sim 1$\%  with respect 
to a nonrelativistic approach, and relativistic corrections to the mean
electron-scattering optical depth are even smaller.
Thus I-fronts, while an intrinsically relativistic phenomenon,
can be accurately treated nonrelativistically in studies of
global cosmological reionization, but not for the isolated H II regions
of short-lived luminous sources at late times, close to the epoch of overlap.

Finally, we note that the original nonrelativistic, cosmological I-front
solutions of \citet{1987ApJ...321L.107S} correctly relate the observed 
line-of-sight distance of the I-front from the source which created it to the 
source luminosity and IGM neutral density, even for cases in which those 
nonrelativistic solutions yield a superluminal expansion speed.  Thus, the
interpretation of the line-of-sight apparent sizes of the cosmological 
H II regions recently observed in the IGM surrounding high-redshift sources
like the SDSS quasars at $z\geq 6$ does not require the relativistic correction
described here.  This somewhat surprising result was noted previously by 
\citet{2003AJ.126..1W} and others \citep[e.g.][]{2005ApJ...620...31Y}.  Our 
results help to illuminate the
connection between the nonrelativistic and relativistic treatments, however,
and distinguish the circumstances under which the relativistic corrections 
{\em do} make a difference.  For example, when we observe a distant
source and match the apparent size of its H~II region along the line
of sight to the {\em nonrelativistic} solution evaluated at some time $t$
(as measured from the source turn-on time), the apparent {\em
transverse} size at the distance of the source will then be given by the
{\em relativistic} solution evaluated at this same time $t$.

\acknowledgements

ITI would like to express his sincere thanks for the hospitality shown
to him by the Kavli Institute for Theoretical Physics, during part of this
research. ES was supported by the National Science
Foundation under  grant PHY99-07949. This work was partially supported
by NASA Astrophysical Theory Program grants NAG5-10825  and  NNG04G177G and 
Texas Advanced Research
Program grant 3658-0624-1999 to PRS. MAA is grateful for the support of a
Department of Energy Graduate Fellowship in Computational Sciences.

\end{document}